\documentclass[fleqn,usenatbib]{mnras}
\usepackage{newtxtext,newtxmath}
\usepackage[T1]{fontenc}
\DeclareRobustCommand{\VAN}[3]{#2}
\let\VANthebibliography\thebibliography
\def\thebibliography{\DeclareRobustCommand{\VAN}[3]{##3}\VANthebibliography}

\usepackage{graphicx}	% Including figure files
\usepackage{amsmath}	% Advanced maths commands
\usepackage{xcolor}
\usepackage{tablefootnote}
\usepackage{orcidlink}

\title[UV emission line properties from $z=1.5$ to $z=4.0$]{No redshift evolution in the rest-frame UV emission line properties of quasars from $z=1.5$ to $z=4.0$} %For now....

\author[M. Stepney et al.]{Matthew Stepney,$^{\orcidlink{0000-0002-7711-0537}\,}$$^{1}$\thanks{E-mail: ms10g17@soton.ac.uk} Manda Banerji,$^{\orcidlink{0000-0002-0639-5141}\,}$$^{1}$ Paul C. Hewett,$^{\orcidlink{0000-0002-6528-1937}\,}$$^{2}$ Matthew J. Temple,$^{\orcidlink{0000-0001-8433-550X}\,}$$^{3}$ Amy L. Rankine,$^{\orcidlink{0000-0002-2091-1966}\,}$$^{4}$
\newauthor
James H. Matthews$^{\orcidlink{0000-0002-3493-7737}\,}$$^{5}$ and Gordon T. Richards$^{\orcidlink{0000-0002-1061-1804}\,}$$^{6}$
\\
    $^{1}$School of Physics and Astronomy, University of Southampton, Southampton, SO17 1BJ, UK\\
    $^{2}$Institute of Astronomy, University of Cambridge, Madingley Road, Cambridge CB3 0HA, UK\\
    $^{3}$Instituto de Estudios Astrof\'{\i}sicos, Universidad Diego Portales, Av. Ej\'ercito Libertador 441, Santiago 8370191, Chile\\
    $^{4}$Institute for Astronomy, University of Edinburgh, Royal Observatory, Blackford Hill, Edinburgh EH9 3HJ, UK\\
    $^{5}$Department of Physics, Astrophysics, University of Oxford, Denys Wilkinson Building, Keble Road, Oxford OX1 3RH, UK\\
    $^{6}$Department of Physics, Drexel University, 3141 Chestnut Street, Philadelphia, PA 19104, USA\\
}

\date{Accepted 2023 July 5. Received 2023 June 6 ; in original form 2023 March 23}

\pubyear{2023}

\begin{document}
\label{firstpage}
\pagerange{\pageref{firstpage}--\pageref{lastpage}}
\maketitle

\begin{abstract}

\noindent We analyse the rest-frame UV spectra of 2,531 high-redshift ($3.5<z<4.0$) quasars from the Sloan Digital Sky Survey DR16Q catalogue. In combination with previous work, we study the redshift evolution of the rest-frame UV line properties across the entire redshift range, $1.5<z<4.0$. We improve the systemic redshift estimates at $z>3.5$ using a cross-correlation algorithm that employs high signal-to-noise template spectra spanning the full range in UV emission line properties. We then quantify the evolution of \ion{C}{iv} and \ion{He}{ii} emission line properties with redshift. The increase in \ion{C}{iv} blueshifts with cosmological redshift can be fully explained by the higher luminosities of quasars observed at high redshifts. We recover broadly similar trends between the \ion{He}{ii} EW and \ion{C}{iv} blueshift at both $1.5<z<2.65$ and $3.5<z<4.0$ suggesting that the blueshift depends systematically on the spectral energy density (SED) of the quasar and there is no evolution in the SED over the redshift range $1.5<z<4.0$. \ion{C}{iv} blueshifts are highest when L/L$_\textnormal{Edd} \geq$ 0.2 and M\textsc{bh} $\geq 10^9 M_\odot$ for the entire $1.5<z<4.0$ sample. We find that luminosity matching samples as a means to explore the evolution of their rest-frame UV emission line properties is only viable if the samples are also matched in the M\textsc{bh} - L/L$_\textnormal{Edd}$ plane. Quasars at $z\geq6$ are on average less massive and have higher Eddington-scaled accretion rates than their luminosity-matched counterparts at $1.5<z<4.0$, which could explain the observed evolution in their UV line properties.

\end{abstract}

\begin{keywords}
quasars: general – quasars: emission lines – line: profiles %Provisional
\end{keywords}

%%%%%%%%%%%%%%%%%%%%%%%%%%%%%%%%%%%%%%%%%%%%%%%%%%

%%%%%%%%%%%%%%%%% BODY OF PAPER %%%%%%%%%%%%%%%%%%

\section{Introduction}

The increase in the size of large spectroscopic samples of high-redshift ($z>1.5$) quasars, with rest-frame ultra-violet (UV) spectra has been rapid over the past decade. The Sloan Digital Sky Survey (SDSS) DR7 catalogue \citep{2010AJ....139.2360S} enabled the study of $\sim 105,000$ rest-frame UV to optical quasar spectra, from which \citet{Shen_2011} constructed a comprehensive catalogue of line properties. With the subsequent SDSS data releases and the introduction of the BOSS spectrograph, the SDSS DR16Q catalogue now contains over 750,000 quasars \citep{2020ApJS..250....8L}, marking a near order of magnitude increase from DR7. Looking to ongoing and upcoming quasar surveys, these numbers are continuing to rapidly increase and push to less optically luminous populations, as well as increasing the numbers at high-redshifts - e.g. the Dark Energy Spectroscopic Instrument (DESI; \citealt{2022arXiv220808517A}), which has an expected quasar target density of $\gtrsim5\times$ that in SDSS DR7. 

Thanks to these large spectroscopic data-sets, we are now able to conduct statistical studies of the rest-frame UV spectra, explore the diversity in their emission line properties and study correlations between the continuum and various emission lines (e.g. \citealt{2007ApJ...666..757S, Richards_2011, 2020MNRAS.492.4553R, Rakshit_2020, Brodzeller_2022, 2022ApJS..263...42W, Rivera_2022}). Characterising the emission line properties also enables single-epoch virial black hole mass estimates for large samples, allowing an exploration of how the UV spectral properties connect to fundamental physical properties of the quasars such as bolometric luminosity, black hole mass and Eddington-scaled accretion rate \citep[e.g,][]{2023MNRAS.523..646T}. The widely observed blueshifts seen in the \ion{C}{iv}$\lambda\lambda1548,1550$ emission lines of high-redshift, high-luminosity quasars are often interpreted as a signature of outflowing gas in the quasar broad line region (BLR; \citealt{2015MNRAS.449.1593B}). Statistical studies of UV quasar spectra then enable these outflows to be linked to other quasar properties. For example, the observed anti-correlation between the \ion{C}{iv} blueshift and \ion{He}{ii} $\lambda1640$ equivalent width (EW) \citep{10.1093/mnras/stt2230,2020MNRAS.492.4553R} can be interpreted as a link between the spectral energy density (SED) of the quasar and BLR outflow velocity \citep{2023MNRAS.523..646T}. 

In parallel to these developments in characterising the demographics of quasars at ``cosmic noon'' ($1.5\lesssim z \lesssim 3.5$), the number of quasars at the highest redshifts ($z\gtrsim5$) has also grown considerably in the last decade \citep[e.g,][]{2016ApJS..227...11B,2016ApJ...833..222J,2019ApJ...884...30W, Fan_2022_review}. A sizeable subset of these also have single-epoch black hole masses measured from rest-frame optical spectra (e.g. \citealt{DeRosa_2014, Mazzucchelli_2017, Schindler_2020, 2022ApJ...941..106F}). These high-redshift quasars are now being targeted as part of ongoing wide-field spectroscopic surveys such as DESI, which has recently confirmed $\sim400$ new quasars at $4.7<z<6.6$ \citep{2023arXiv230201777Y}. These numbers are only expected to increase further with new observations from the 4-metre Multi-Object Spectroscopic Telescope (4MOST; \citealt{2019Msngr.175...42M}).

The discovery of the first quasar at $z>7$ \citep{2011Natur.474..616M} showed that its rest-frame UV spectrum was very similar to lower redshift quasars, of similar luminosity, but that the \ion{C}{iv} blueshift was significantly larger than the other known $z>2$ quasars. With larger samples becoming available at high-redshifts, there have been suggestions that high-redshift quasars might in general display stronger \ion{C}{iv} blueshifts \citep[e.g,][]{2019MNRAS.487.3305M,Schindler_2020}, as well as higher broad absorption line fractions and velocities \citep[e.g,][]{2022Natur.605..244B,2023arXiv230109731B}. However, these samples are still small enough that these results might be affected by limited sample statistics \citep[e.g,][]{2019MNRAS.487.1874R,2021ApJ...923..262Y}. While there is an indication that there could be a real evolution in the rest-frame UV line properties driven by fundamental differences in the super massive black hole (SMBH) population at the highest redshifts, neither the line properties nor the physical properties of quasars have been measured consistently across the entire range in redshift ($1.5\lesssim z \lesssim7.0$).
 
In the radiation line-driven disc wind paradigm, ionised gas opaque to the UV continuum photons emitted by the accretion disc, due to UV line transitions, trace the streamlines of outflows and hence the \ion{C}{iv} blueshift becomes a diagnostic for the outflow velocity. By implication, the observed increase in \ion{C}{iv} blueshift in the highest redshift quasars at $z\gtrsim6$ suggests that quasar winds are potentially evolving very rapidly on time-scales of $\sim200-300$ Myrs \citep{2019MNRAS.487.3305M}. Thus far, the statistical studies of the rest-frame UV line properties, of quasars, have mostly focused on lower redshifts where we have samples of several hundred thousand spectra. In particular, below z=2.65 the $\ion{Mg}{ii}\,\lambda2800$ emission line is present in the SDSS spectrum, which enables the UV line properties to be linked to black hole mass. However, in lieu of the results at the highest redshifts, it is worth extending the studies of the evolution of the rest-frame UV spectra by even a few hundred Myr while also ensuring sufficient sample statistics. We therefore focus on quasars at redshifts $3.5<z<4.0$ in this paper. 

There are thousands of SDSS quasar spectra at $3.5<z<4.0$ allowing us to study the distribution of rest-frame UV line properties without the need to rely solely on sample averages. However, as accurate \ion{C}{iv} blueshift measurements are key to this work, we also need to ensure that accurate systemic redshifts estimates are available for the full sample. At $z<4.0$, the \ion{C}{iii]} complex is present in the SDSS observed wavelength range, which we can use to improve the systemic redshift estimates available from the SDSS pipeline, and which then allows us to measure the \ion{C}{iv} blueshift based on the UV spectrum alone. This is crucial because while rest-frame optical emission lines or, even better, host galaxy ISM lines are the gold standard for systemic redshifts, they require either NIR spectroscopic follow-up or ALMA follow-up, which is not feasible for samples of thousands of quasars. 

An additional aim of our work is to probe lower signal-to-noise SDSS spectra than what is currently analysed at $z\sim2$. The median continuum signal-to-noise of the \citet{2020MNRAS.492.4553R} sample is $\sim6.4$, compared to $\sim4.9$ at $3.5<z<4.0$. Ensuring our techniques are robust at these signal-to-noise ratios is important for the study of high redshift quasar spectra in SDSS. At $z\gtrsim2.7$ the SDSS wavelength coverage also means the \ion{Mg}{ii} emission line, widely used as a reliable single-epoch virial black hole mass estimator, is redshifted out of the observed spectrum, and hence we must turn to \ion{C}{iv} for SMBH masses. A further aim of this work is to determine whether changing the SMBH mass estimator, compared to e.g. the $1.5<z<2.65$ samples, has an impact on the observed trends in UV line properties with SMBH mass and Eddington-scaled accretion rate \citep[e.g,][]{2023MNRAS.523..646T}. By measuring robust blueshifts from only the rest-frame UV spectra, and doing so at modest signal-to-noise, as well as demonstrating that the trends in these line properties are not sensitive to the SMBH mass estimator, we potentially open the door to studies of the evolution in UV line properties for much larger statistical samples out to high-redshifts e.g. those being assembled with DESI and 4MOST. 

The structure of the paper is as follows. In Section \ref{sec:data} we present the selection of the $3.5<z<4.0$ sample and its subsequent post-processing. Section \ref{sec:Methods} describes the spectral reconstruction recipe used to characterise the emission line properties. Our key results are compared to the \citet{2020MNRAS.492.4553R,2023MNRAS.523..646T} $1.5<z<3.5$ samples in Section \ref{sec:Results}, before discussing their implications in the context of the redshift evolution of quasar-driven outflows in Section \ref{sec:discussion}. Vacuum wavelengths are employed throughout the paper and we adopt a $\Lambda \textnormal{CDM}$ cosmology with $h_0$ = 0.71, $\Omega_\textnormal{M}$ = 0.27 and $\Omega_\Lambda$ = 0.73 when calculating quantities such as quasar luminosities.

\section{Data} \label{sec:data}

Our quasar sample is drawn from the final installment of the Sloan Digital Sky Survey IV (SDSS-IV) quasar catalogue, otherwise referred to as the sixteenth data release of the extended Baryon Oscillation Spectroscopic Survey \citep[eBOSS;][]{Dawson_2016}. The catalogue, which we will refer to as DR16Q, is  comprised of 750,414 quasars, including the $\sim$526,000 known quasars from SDSS-I/II/III and a further 225,082 quasars new to SDSS catalogues \citep{2020ApJS..250....8L}. We select a sample of 2775 non-BAL quasars (BAL probability$\leq$0.7) in the redshift range $3.5<z<4.0$ from the DR16Q quasar catalogue, which have a median signal-to-noise $\geq3.0$ per pixel in the rest-frame interval 1265-1950\AA. Above $z>4$ the numbers of quasars in DR16Q are relatively small precluding a statistical analysis of their UV line demographics. Furthermore at $z>4$ the \ion{C}{iii]} emission line complex redshifts out of the SDSS observed wavelength range, which means systemic redshift estimates based solely on rest-frame UV spectra, becomes increasingly unreliable \citep{10.1111/j.1365-2966.2010.16648.x, 2022ApJS..263...42W}. We therefore restrict our sample to $3.5<z<4.0$. 

We follow a routine similar to that described by \citet{10.1111/j.1365-2966.2010.16648.x} and later adopted by \citet{2016MNRAS.461..647C, 2017MNRAS.465.2120C} and \citet{2020MNRAS.492.4553R} to remove narrow absorption features and sky line residuals from the spectra. First, we define a 61-pixel median filtered \texttt{pseudo-continuum}. We then exclude any pixels that fall within 6\AA\, of the strong night-sky emission lines at 5578.5\AA\, and 6301.5\AA\,. Pixels that fall below $2\sigma$ of the \texttt{pseudo-continuum} are regarded as narrow line absorption features and hence removed from the spectrum with a grow radius of two pixels. The removed pixels are consequently replaced by their corresponding pixel in the \texttt{pseudo-continuum} spectrum. 

\subsection{Improving SDSS systemic redshift estimates} \label{sec:redshift estimates}

The accuracy to which the SDSS pipeline estimates the systemic redshift of quasars has greatly improved since DR7 \citep{Shen_2011}. The DR16Q catalogue includes a number of redshift estimates for each quasar \citep{2020ApJS..250....8L} e.g. automated classifications from the SDSS spectroscopic pipeline, visual inspection estimates (which reveal that only $2.1\%$ of the automated classifications result in a catastrophic failure) and redshifts based on principal component analysis (PCA) \citep{2020ApJS..250....8L}. In this work, we define $Z_{\textnormal{SDSS}}$ as the ``primary'' redshift presented in DR16Q, defined by either visual inspection or automated classifications. For our sample, $Z_{\textnormal{SDSS}} = Z_{\textnormal{VI}}$ for 2,527 quasars and $Z_{\textnormal{SDSS}} = Z_{\textnormal{AUTO}}$ for the remaining 4 objects. Nevertheless, these redshift estimates do not as yet account for the systematic velocity offsets between different quasar emission lines as a function of the UV emission line morphologies. This becomes increasingly true at high redshifts, where the \ion{C}{iv} emission line begins to dominate the redshift estimates.

\begin{figure*}
 \includegraphics[width=\linewidth]{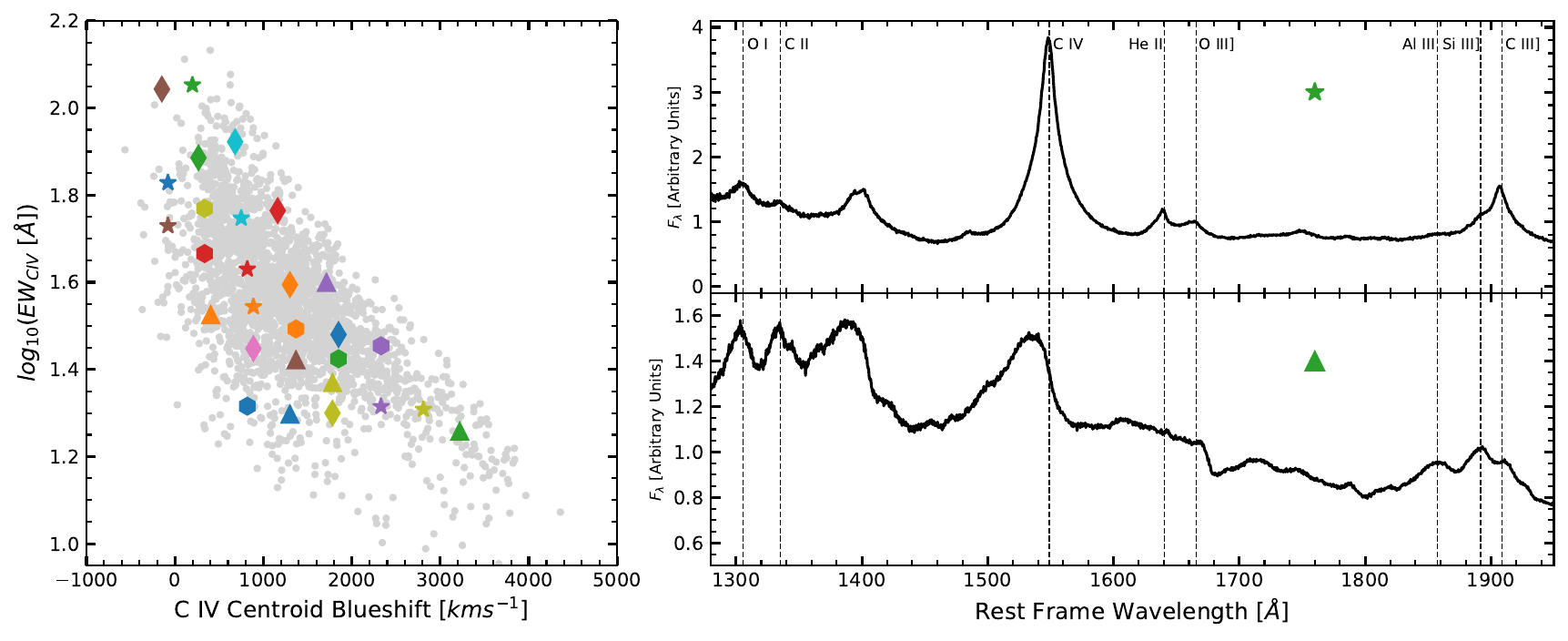} 
  \caption{The left-hand panel shows the \ion{C}{iv} blueshift versus equivalent width for the $3.5<z<4.0$ quasar sample with the locations of the 33 high signal-to-noise templates used in the cross-correlation algorithm to calculate systemic redshifts. The spectra for templates at the extremes of the \ion{C}{iv} emission space are also presented on the right. The high \ion{C}{iv} EW and low \ion{C}{iv} blueshift template (top right) features a strong symmetric \ion{C}{iv} line profile. Conversely, the low \ion{C}{iv} EW and high \ion{C}{iv} blueshift template (bottom right) features both a weak and asymmetric \ion{C}{iv} line profile and stronger \ion{Si}{iii]} relative to \ion{C}{iii]}. Accounting for these systematic changes in the SEDs as a function of the \ion{C}{iv} line properties is critical to producing accurate systemic redshifts.}
 \label{fig:DR16_Templates}
\end{figure*}

Accurate systemic redshifts are crucial to our study and we therefore calculate updated systemic redshifts for our sample of high-redshift quasars as follows. A visual inspection of all 2775 $3.5<z<4.0$ non-BAL quasar spectra shows that 244 quasars do not show clear broad emission line features and we therefore consider these objects to be misclassified. This number represents a larger fraction of misclassified objects than those found in among all SDSS quasar targets \citep{2020ApJS..250....8L}. To improve the systemic redshift estimates, we adopt a filtering and `cross-correlation’ scheme as described in section 4.2 of \citet{10.1111/j.1365-2966.2010.16648.x}. There are two key differences compared to \citet{10.1111/j.1365-2966.2010.16648.x}. First, we employ 33 high signal-to-noise composite spectra. The composite spectra are generated using median stacks of individual SDSS quasar spectra whose systemic redshifts are calculated using the 1600-3000\AA\, wavelength region, including the \ion{C}{iii]} emission complex and the \ion{Mg}{ii} emission line. The 33 composite spectra span the full range of  \ion{C}{iv} emission morphologies in equivalent-width and blueshift space, as illustrated in Fig. \ref{fig:DR16_Templates}. The properties of the composite spectra are summarised in Table \ref{tab:HSN_TEMPLATES}. Second, the cross-correlation between each template and an individual quasar spectrum is performed with the quasar redshift as a free parameter. Finally, we create a bespoke template from the weighted mean of the seven composite spectra with the largest cross-correlation values, using the cross-correlation coefficients as the weights, and perform one last cross-correlation with the individual spectrum, again with the redshift as a free parameter. 

\begin{table*}
    \centering
    \caption{We present estimates of the average signal-to-noise, \ion{C}{iv} EW and \ion{C}{iv} blueshift for the 33 high signal-to-noise composite spectra used to correct the systemic redshifts of our sample. The signal-to-noise estimates are calculated assuming the individual spectra have a median continuum signal-to-noise $\sim5$. We also provide filenames for the composites, which are available as online only supplementary material, in addition to the total number of spectra used to construct them. 62,464 quasar spectra were used in total.}
    
    \begin{tabular}{|l|c|c|c|c|}
         \hline
         \hline
         Filename & No. of Spectra & Signal-to-noise & \ion{C}{iv} EW [\AA] & \ion{C}{iv} blueshift [$\textnormal{km\,s}^{-1}$] \\
         \hline
         \hline
         mc$\_$00000500$\_$hew500.fits & 500 & 110 & 113 & 197 \\
         mc$\_$00000500$\_$lew500p1.fits & 5460 & 370 & 33.5 & 404 \\
         mc$\_$00000500$\_$lew500p2.fits & 5461 & 370 & 46.3 & 336 \\
         mc$\_$00000500$\_$lew500p3.fits & 5461 & 370 & 58.8 & 336\\
         mc$\_$00000500$\_$lew500p4.fits & 5459 & 370 & 76.8 & 266 \\
         mc$\_$00000500$\_$lew500.fits & 500 & 110 & 21.7 & 404 \\
         mc$\_$05001000$\_$hew500.fits & 500 & 110 & 83.6 & 680 \\
         mc$\_$05001000$\_$lew500p1.fits & 3942 & 310 & 28.1 & 887 \\
         mc$\_$05001000$\_$lew500p2.fits & 3942 & 310 & 35.0 & 887 \\
         mc$\_$05001000$\_$lew500p3.fits & 3942 & 310 & 42.7 & 818\\
         mc$\_$05001000$\_$lew500p4.fits & 3942 & 310 & 55.9 & 749 \\
         mc$\_$05001000$\_$lew500.fits & 500 & 110 & 20.7 & 818 \\
         mc$\_$10001500$\_$hew500.fits & 500 & 110 & 58.1 & 1162 \\
         mc$\_$10001500$\_$lew500p1.fits & 2402 & 250 & 26.4 & 1369 \\
         mc$\_$10001500$\_$lew500p2.fits & 2403 & 250 & 31.2 & 1369\\
         mc$\_$10001500$\_$lew500p3.fits & 2403 & 250 & 39.3 & 1300\\
         mc$\_$10001500$\_$lew500.fits & 500 & 110 & 19.8 & 1300 \\
         mc$\_$15002000$\_$hew500.fits & 500 & 110 & 39.8 & 1712 \\
         mc$\_$15002000$\_$lew500p1.fits & 841 & 150 & 23.4 & 1781 \\
         mc$\_$15002000$\_$lew500p2.fits & 840 & 150 & 26.6 & 1850 \\
         mc$\_$15002000$\_$lew500p3.fits & 842 & 150 & 30.2 & 1850 \\
         mc$\_$15002000$\_$lew500.fits& 500 & 110 & 19.9 & 1781 \\
         mc$\_$20002500$\_$010025.fits& 799 & 140 & 20.6 & 2330 \\
         mc$\_$20002500$\_$025050.fits & 674 & 130 & 28.4 & 2330\\
         mc$\_$25003000$\_$all.fits & 744 & 140 & 20.3 & 2809 \\
         mc$\_$30006000$\_$all.fits & 577 & 120 & 18.1 & 3219 \\
         mc$\_$m05000000$\_$hew500.fits & 500 & 110 & 111 & -148\\
         mc$\_$m05000000$\_$lew500p1.fits & 1647 & 200 & 39.9 & -78.8 \\
         mc$\_$m05000000$\_$lew500p2.fits& 1647 & 200 & 53.7 & -78.8 \\
         mc$\_$m05000000$\_$lew500p3.fits & 1647 & 200& 67.4 & -78.8\\
         mc$\_$m05000000$\_$lew500p4.fits & 1647 & 200 & 80.4 & -148\\
         mc$\_$m05000000$\_$lew500.fits & 500 & 110 & 27.1 & -148 \\
         mc$\_$m10000500$\_$all.fits & 742 & 140 & 54.9 & -562 \\
         \hline
         \hline
    \end{tabular}
    \label{tab:HSN_TEMPLATES}
\end{table*}

\begin{figure*}
 \includegraphics[width=\linewidth]{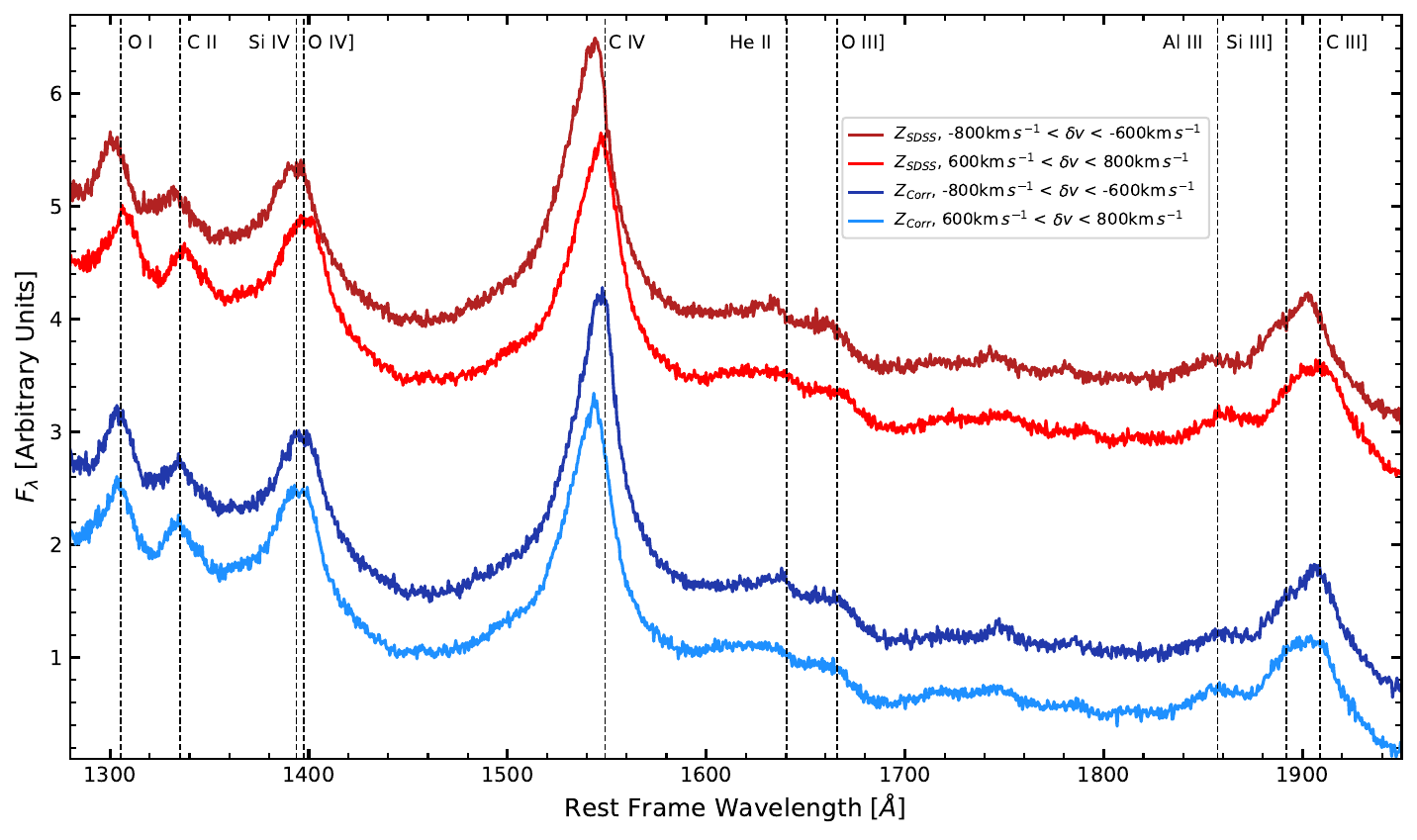} 
 
  \caption{Composites of quasar spectra with redshift corrections -800\,$\textnormal{km\,s}^{-1}$ < $\delta v$ < -600\,$\textnormal{km\,s}^{-1}$ or 600\,$\textnormal{km\,s}^{-1}$ < $\delta v$ < 800\,$\textnormal{km\,s}^{-1}$. We present composites before a redshift correction is applied (red) and afterwards (blue). The corrected composites (blue) align with the lower ionisation lines, such as \ion{O}{i}, \ion{C}{ii}, \ion{Si}{iv]} and \ion{O}{iv} far more convincingly than when the original SDSS systemic redshifts are used.}
 \label{fig:ZHEWETT}
\end{figure*}

\autoref{fig:DR16_Templates} (right) depicts two templates from the top-left and bottom-right of the \ion{C}{iv} emission space shown in Fig. \ref{fig:DR16_Templates} (left). The main difference in our approach to the estimation of systemic redshifts employing the \ion{C}{iv} emission line is the use of templates that span the full range of emission line morphologies. The SDSS scheme employing their principal component analysis five-components, for example, does not have the capability to reproduce the full range of quasar UV SEDs, particularly at the extremes of the CIV emission-line space. We present composite spectra for objects with systemic redshift corrections -800\,$\textnormal{km\,s}^{-1}$ < $\delta v$ < -600\,$\textnormal{km\,s}^{-1}$ or 600\,$\textnormal{km\,s}^{-1}$ < $\delta v$ < 800\,$\textnormal{km\,s}^{-1}$ in Fig. \ref{fig:ZHEWETT}. \autoref{fig:ZHEWETT} illustrates that the systemic redshifts have dramatically improved with respect to the initial SDSS estimates. The centroids of the semi-forbidden $\ion{Si}{iii]}\lambda1892$ and $\ion{C}{iii]}\lambda 1908$ emission are more consistent with the rest-frame wavelengths, after a correction to the systemic redshift was applied (Fig. \ref{fig:ZHEWETT} blue). This is in contrast with the \ion{C}{iv} emission since semi-forbidden lines are produced at larger radii and should therefore not be present in outflows. We also note a similar effect in the higher energy $\ion{O}{i]}$ $\lambda1306$, $\ion{C}{ii]}$ $\lambda1335$ and $\ion{Si}{iV]}$  $\lambda1398$ emission lines as well as the $\ion{He}{ii}$ $\lambda1640$ and $\ion{O}{iii]}$ $\lambda1665$ complex. 

We compare the systemic redshifts presented in DR16Q against the corrected redshifts, from this work, and find that 781 objects require systemic redshift corrections in excess of 500 $\textnormal{km\,s}^{-1}$, 244 of which are in excess of 1000 $\textnormal{km\,s}^{-1}$, corresponding to $\sim31$ per cent and $\sim10$ per cent of the sample, respectively. 

\section{Methods} \label{sec:Methods}

\subsection{Spectral Reconstructions using Mean Field Independent Component Analysis (MFICA)}

\begin{figure*}
 \includegraphics[width=\linewidth]{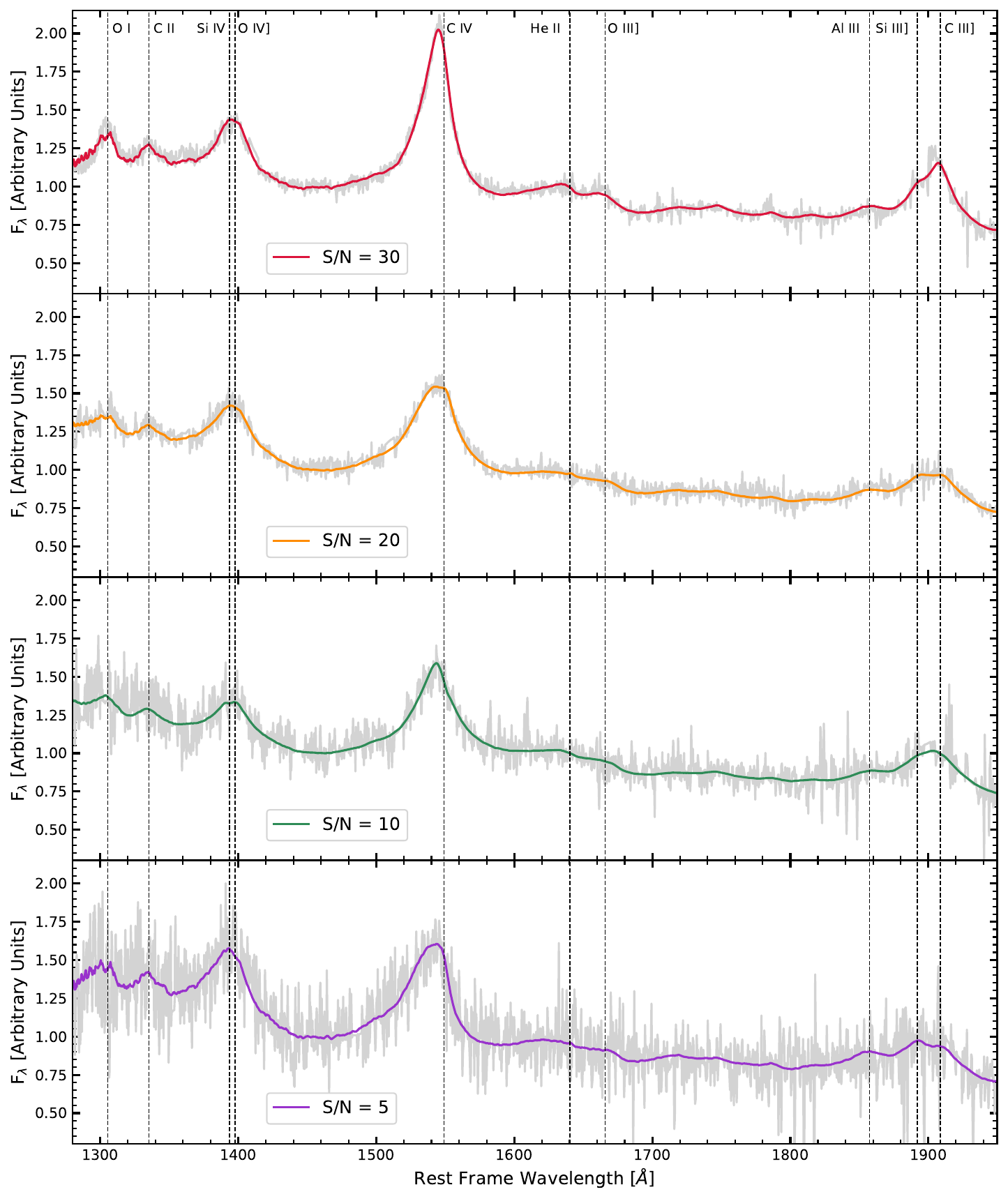} 
  \caption{Example quasar spectra from our sample at $3.5<z<4.0$ with signal-to-noise of 30 (top), 20 (top-middle), 10 (bottom-middle) and 5 (bottom), with their corresponding mean-field independent component analysis reconstructions overlaid. We can see that the MFICA reconstructions have the effect of boosting the signal-to-noise of the SDSS spectra, and hence robust UV emission line analysis becomes achievable.}
 \label{fig:Spectra_SN}
\end{figure*}

An objective of our work is to compare the emission line properties of the $3.5<z<4.0$ quasars to lower redshift quasars at $1.5<z<3.5$ using exactly the same methods to analyse and fit the quasar spectra. This avoids biases in the measured line properties due to the different line-fitting procedures employed for quasars at different redshifts. 

Our lower redshift comparison sample of quasars used throughout this work is drawn from \citet{2020MNRAS.492.4553R} and \citet{2023MNRAS.523..646T}, covering the redshift range $1.5<z<3.5$. The rest-frame UV line demographics for quasars with $1.5<z<2.65$ and where the \ion{Mg}{ii} emission line is available for robust SMBH mass measurements, has been studied in detail by \citet{2023MNRAS.523..646T}. Analogous to \citet{2020MNRAS.492.4553R} we therefore employ a technique called Mean-Field Independent Component Analysis (MFICA; \citealt{6789937}) to provide essentially noise-free reconstructions of each observed spectrum through a linear combination of 7-10 component spectra (see section 4.2 of \citealt{2020MNRAS.492.4553R} for more details). The component spectra are the same as those used by \citet{2020MNRAS.492.4553R} at $1.5<z<3.5$. They were generated using a sample of $\sim4000$ $1.5<z<3.5$ spectra at intermediate \ion{C}{iv} EWs, 20-40\AA\,, and a further two samples of $\sim2000$ $1.5<z<3.5$ spectra with \ion{C}{iv} EW $>40$\AA\, and $<20$\AA. The scheme for deriving MFICA-components for use on astronomical data is described in \citet{2013MNRAS.430.3510A}. Three sets of components were necessary to account for the extensive array of \ion{C}{iv} emission line morphologies seen in the sample. For a visualisation of the three sets of MFICA components, which will be made available as online only supplementary material, see Appendix \ref{Appendix:MFICA}. The MFICA components are highly effective for our purpose, however, the exact form of the components is not important. Rather, the key requirement is that the components can be combined to produce accurate reconstructions of the quasar spectra. Alternative approaches using principal component analysis or non-negative matrix factorization should also work well. 

Spectrophotometric calibration effects introduce both a blue excess and red decrement to the BOSS spectra \citep{Dawson_2012,2012A&A...548A..66P}. Consequently, we observe a $\pm10\%$ multiplicative factor in the spectrophotometry as well as an additional effect introduced by the varying dust extinction towards each quasar. Since we aim to reconstruct the quasar spectra with linear combinations of fixed components, any wavelength dependent multiplicative factor needs to be removed, hence we implement a "morphing" recipe to standardise the shape of the quasar spectra. For a full description of the morphing recipe see section 4.1 of \citet{2020MNRAS.492.4553R}. The reference quasar SED used to "morph" the quasar spectra is a model quasar spectrum, discussed extensively in \citet{2012MNRAS.424.2876M} and \citet{2021MNRAS.508..737T}\footnote{\url{https://github.com/MJTemple/qsogen}}. 

We use the \textsc{emcee} Python package\footnote{\url{https://github.com/dfm/emcee}} \citep{emcee} for the MFICA spectral reconstructions. The \textsc{emcee} package is a Python implementation of the affine-invariant ensemble sampler for Markov-Chain Monte Carlo (MCMC) simulations, proposed by \citet{2010CAMCS...5...65G}. We permit the \textsc{emcee} package to explore an N-Dimensional, Gaussian likelihood function, where N represents the number of MFICA components (N=7 for the high \ion{C}{iv} EW set and N=10 otherwise), and apply uniform priors to the component weights. Example spectra, with various signal-to-noise characteristics are presented in Fig. \ref{fig:Spectra_SN} together with their MFICA reconstructions. A key conclusion is that the $1.5<z<3.5$ MFICA components can successfully reconstruct the $3.5<z<4.0$ spectra with a range of signal-to-noise ratios. The MFICA has the effect of boosting the signal-to-noise characteristics of the SDSS spectra affording us the opportunity to analyse weaker UV emission lines such as \ion{He}{ii}. We have checked that different realisations of the spectra perturbed consistent with the Gaussian noise on each spectraum, result in essentially identical MFICA reconstructions. Although the component weights themselves change, the line properties derived from the reconstructions are therefore unchanged by the noise on each spectrum. A further advantage of MFICA over parametric fitting of emission lines is the ability to accurately reproduce asymmetric emission line features. For a full discussion, and comparison to catalogues where line properties are determined via parametric fitting, refer to Appendix \ref{Appendix:Shen}. 

\subsection{Line properties} 

A central aim of this work is to assess whether there is any redshift evolution in the rest-frame UV emission line properties from $1.5<z<4.0$. To determine the \ion{C}{iv} EW and blueshift, we first define a power-law continuum, $f(\lambda) \propto \lambda^{-\alpha}$. We then follow the non-parametric approach discussed by \citet{2016MNRAS.461..647C,2017MNRAS.465.2120C}, whereby the median values of F$_{\lambda}$ in the two wavelength regions 1445–1465\AA\, and 1700–1705\AA\, are used to determine the power-law approximation. Then, \ion{C}{iv} EW measurements are made via numerical integration. Due to the well known asymmetry of the \ion{C}{iv} emission line \citep{Richards_2011}, we define the blueshifts as the difference between the line centroid and rest-frame wavelength: 

\begin{equation}
    \ion{C}{iv} \, \, \textnormal{blueshift} = c(\lambda_r -\lambda_{\textnormal{half}})/\lambda_r \, \, \,[\textnormal{km\,s}^{-1}]
    \label{eq:blueshift}
\end{equation}

\noindent where c is the velocity of light, $\lambda_{\textnormal{half}}$ is the rest-frame wavelength of the flux-weighted line centroid and $\lambda_r$ is 1549.48\AA\ for the \ion{C}{iv} doublet.
We note that several papers in the literature exploring the UV emission line properties of high-redshift quasars (e.g. \citealt{2019MNRAS.487.3305M,Schindler_2020}) use the observed wavelength corresponding to the peak of the line to define the \ion{C}{iv} blueshift. This is often necessary at low signal-to-noise but does not fully capture the often significant flux in the blue wing of the \ion{C}{iv} emission line.

\subsection{Luminosities and black hole masses} \label{SEC:PQP}

We infer rest-frame monochromatic continuum luminosities $\lambda L_\lambda$ at $\lambda=3000$ and 1350\,\AA\ (hereafter $L_{3000}$ and $L_{1350}$, respectively) using the PSF magnitudes from SDSS photometry reported in the \citet{2020ApJS..250....8L} DR16Q catalogue.
The SDSS photometry is corrected for Galactic dust extinction using the reddening law presented by \citet{2011ApJ...737..103S} and the quasar-specific pass-band attenuations described in section 3.1.4 of \citet{2021MNRAS.508..737T}.
For each object, we then fit a quasar SED model \citep{2021MNRAS.508..737T} to the extinction-corrected photometry using the improved spectroscopic redshifts described in Section \ref{sec:redshift estimates},
with the quasar luminosity and continuum reddening $E(B-V)$ as the only free parameters. 
The $E(B-V)$ accounts for the (slight) variations observed in ultraviolet continuum slope and allows a robust estimation of the rest-frame $L_{3000}$, which in turn allows us to estimate bolometric luminosities in the same way for our $z>3.5$ and $z\approx2$ samples.
We exclude pass-bands with rest-frame wavelengths $\lambda < 1215$\,\AA, which results in the use of data from the \textit{riz} pass-bands for our $z>3.5$ sample, and the \textit{griz} pass-bands for our $1.5<z<2.65$ comparison sample from \citet{2020MNRAS.492.4553R}.

As the $\ion{Mg}{ii}\,\lambda2800$ emission line is absent from SDSS spectra at redshifts $z\goa2.7$, we estimate SMBH masses for our $3.5<z<4.0$ sample using the FWHM of the \ion{C}{iv} emission. Due to the asymmetric wing, blue-ward of the \ion{C}{iv} peak emission, we apply a correction to the $(\textnormal{FWHM})_{\ion{C}{iv}}$ derived from the correlation between the FWHM of the \ion{C}{iv} and $\textnormal{H}_{\beta}$ emission lines - see Equation 5 and section 4.3 of \citet{2017MNRAS.465.2120C} for further details. 

\begin{equation}
    (\textnormal{FWHM})_{\ion{C}{iv},\textnormal{Corr.}} = \frac{(\textnormal{FWHM})_{\ion{C}{iv}}}{(0.36 \pm 0.03)\left(\frac{\ion{C}{iv} \, \, \textnormal{blueshift}}{10^3 \textnormal{km\,s}^{-1}}\right) + (0.61 \pm 0.04)}
    \label{eq:CIV_Correction}
\end{equation}

\noindent The functional form of Eqn. \ref{eq:CIV_Correction} leads to inappropriate mass estimates for objects with modest (or indeed negative) blueshifts, hence we only apply the correction on those objects whose \ion{C}{iv} blueshift is >500$\textnormal{km\,s}^{-1}$ \citep[Figure 6;][]{2017MNRAS.465.2120C}. We correct the $(\textnormal{FWHM})_{\ion{C}{iv}}$ under the assumption that the \ion{C}{iv} blueshift = 500$\textnormal{km\,s}^{-1}$, for the quasars that fall short of this threshold although the results would be qualitatively unchanged if no correction was applied to quasars with blueshifts of $<$500$\textnormal{km\,s}^{-1}$. The mass estimates are then calculated using Eq. \ref{eq:CIV_MASS}: 

\begin{equation}
    (\textnormal{M}_{\textnormal{bh}})_{\ion{C}{iv},\textnormal{Corr.}} = 10^{6.71} \left(\frac{(\textnormal{FWHM})_{\ion{C}{iv},\textnormal{Corr.}}}{10^3 \textnormal{kms}^{-1}}\right)^2 \left(\frac{\lambda \textnormal{L}_{\lambda}(1350A)}{10^{44} \textnormal{erg} \, \textnormal{s}^{-1}}\right)^{0.53}
    \label{eq:CIV_MASS}
\end{equation}

\noindent To determine the Eddington-scaled accretion rates of our sample, we use the 3000\AA\ rest-frame luminosities and \ion{C}{iv} derived SMBH masses. We apply the bolometric correction $BC_{3000}$ = 5.15 \citep{Shen_2011,2020MNRAS.492.4553R} to convert the 3000\AA\ rest-frame luminosities to bolometric luminosities from which we can calculate the Eddington-scaled accretion rates for our sample. A catalogue of all line properties derived from these spectra will be available as online only supplementary material. Details of the catalogue are presented in Table \ref{tab:Catalogue_structure}.

\begin{table*}
    \centering
    \caption{The format of the table containing the emission line properties from our MFICA reconstructions. The table is available in a machine-readable format in the online journal.}
    \begin{tabular}{|l|l|l|}
         \hline
         \hline
         Header & Units & Description \\
         \hline
         \hline
         SPEC$\_$FILE & \textemdash & SDSS spec file name \\
         RA & Degrees & Right Ascension \\
         DEC & Degrees & Declination \\
         REDSHIFT & [\textemdash,\,\textemdash,\,$\textnormal{km\,s}^{-1}$] & Z$_{\textnormal{SDSS}}$, Z$_{\textnormal{Corrected}}$, Velocity shift \\
         LOG$\_$L1350 & log$_{10}(\textnormal{erg\,s}^{-1}$) & Monochromatic continuum luminosity at 1350\AA \\
         LOG$\_$L3000 & log$_{10}(\textnormal{erg\,s}^{-1}$) & Monochromatic continuum luminosity at 3000\AA \\ 
         LOG$\_$MBH$\_$CIV$\_$COATMAN & log$_{10}$(M$_{\odot}$) & \ion{C}{iv} derived M\textsc{bh}$^a$ \\
         LOG$\_$LAMBDA$\_$EDD & \textemdash & Eddington-scaled accretion rate\\
         C$\_$IV & [\AA,\, $\textnormal{km\,s}^{-1}$,\, $\textnormal{km\,s}^{-1}$,\, $\textnormal{km\,s}^{-1}$] & \ion{C}{iv}: EW, Centroid blueshift, FWHM and FWHM$_{\textnormal{Coat,Corr}}^a$ \\
         HE$\_$II & \AA & \ion{He}{ii}: EW \\
         \hline
         \hline
         \multicolumn{3}{l}{$^a$ The \ion{C}{iv} FWHM \cite{2017MNRAS.465.2120C} correction has been applied.}
    \end{tabular}
    \label{tab:Catalogue_structure}
\end{table*}

\section{Results} \label{sec:Results}

In this section we present the \ion{C}{iv} and \ion{He}{II} line properties for the $3.5<z<4.0$ quasar sample and compare to quasars at $1.5<z<3.5$ from the literature. 

\subsection{\ion{C}{iv} emission}

\begin{figure} 
 \includegraphics[width=\linewidth]{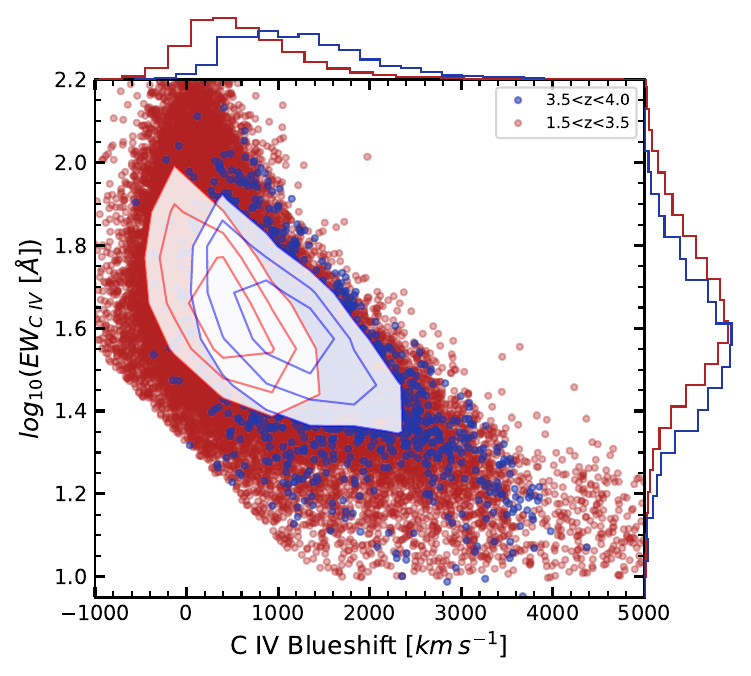}
 \caption{The \ion{C}{iv} emission space for both the $1.5<z<3.5$ \citep[red,][]{2020MNRAS.492.4553R,2023MNRAS.523..646T} and $3.5<z<4.0$ (blue) samples. Density contours encircle  68, 50 and 25 per cent of the sample, respectively. Marginalised distributions of the \ion{C}{iv} blueshift and EW are also shown. While the distributions feature the same characteristic shape, the $3.5<z<4.0$ sample is biased to higher \ion{C}{iv} blueshift and lower \ion{C}{iv} EW. The following criteria was applied to remove poorly fit spectra from both samples; 
 \newline log$_{10}$(\ion{C}{iv} EW [\AA\,]) $< -2.3077 \times 10^{-4} \times$ \ion{C}{iv} blueshift [$\textnormal{km\,s}^{-1}$] $+ 1.3231$.} 
 \label{fig:CIV-Unmatched}
\end{figure}

\autoref{fig:CIV-Unmatched} shows the distribution of the \ion{C}{iv} blueshifts and equivalent widths in the $3.5<z<4.0$ sample as well as the $1.5<z<3.5$ sample from \citet{2020MNRAS.492.4553R} and \citet{2023MNRAS.523..646T}. The \citet{2020MNRAS.492.4553R} sample utilises SDSS DR14 so we supplement the sample with a small number of additional quasars in the same redshift range from DR16Q. Consistent with the findings of \citet{Richards_2011, 2016MNRAS.461..647C, 2017MNRAS.465.2120C, 2020MNRAS.492.4553R}, we observe an anti-correlation between the \ion{C}{iv} blueshift and \ion{C}{iv} EW of the emission lines. Stronger emission lines are generally symmetric and show modest \ion{C}{iv} blueshifts while weaker lines exhibit a range of blueshifts with a clear tail extending to very high blueshifts of several 1000 km/s. While we find that both the $1.5<z<3.5$ and $3.5<z<4.0$ samples have the same characteristic distribution in the \ion{C}{iv} emission space, as expected, the $3.5<z<4.0$ objects do not populate the same region as the $1.5<z<3.5$ sample. Due to the well-known Baldwin effect, the more luminous $3.5<z<4.0$ quasars are biased towards lower \ion{C}{iv} EWs and the $3.5<z<4.0$ quasars also show a tail extending to much higher \ion{C}{iv} blueshifts. 

\subsubsection{Matching the samples by quasar properties}

To ensure a fair comparison between $3.5<z<4.0$ and cosmic noon, we match our $3.5<z<4.0$ quasars to the $1.5<z<2.65$ DR16Q sample from \citet{2023MNRAS.523..646T}, by choosing the nearest $1.5<z<2.65$ quasar to each $3.5<z<4.0$ quasar in a variety of quasar physical properties. Specifically we match in turn by 3000\AA\, continuum luminosity and \textit{both} SMBH mass and Eddington-scaled accretion rate. For the lower redshift sample the SMBH masses are derived from the \ion{Mg}{ii} emission line, while at higher redshifts of $z>2.65$ they are derived instead using \ion{C}{iv}. A key aim of this work is to investigate how changing the SMBH mass estimator potentially affects the dependence of UV line properties on SMBH mass and accretion rate.  

\begin{figure} 
 \includegraphics[width=\linewidth]{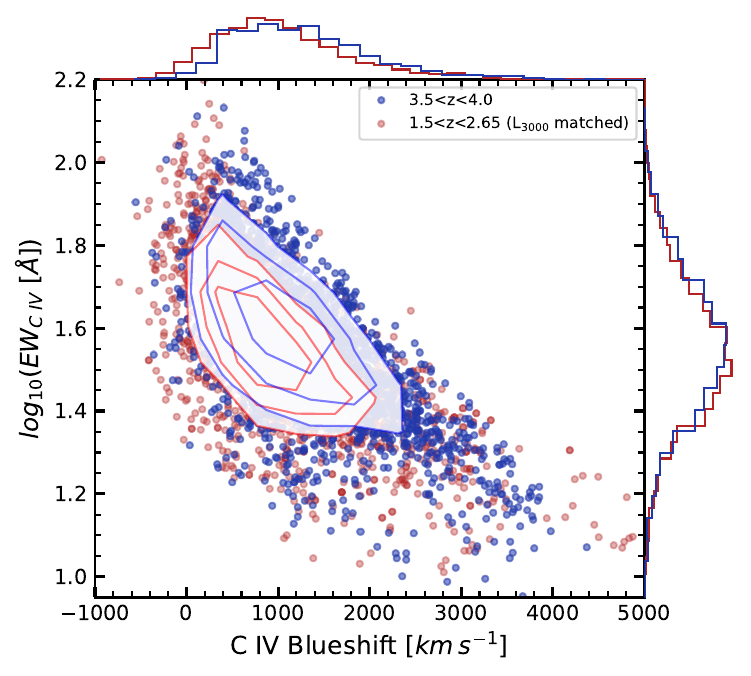}
 \caption{The \ion{C}{iv} emission space for the $3.5<z<4.0$ sample (blue) and the L$_{3000}$ matched $1.5<z<2.65$ quasars (red) from \citet{2023MNRAS.523..646T}. Density contours encircle 68, 50 and 25 per cent of the sample, respectively.}
 \label{fig:CIV-L3000_matched}
\end{figure}

\begin{figure} 
 \includegraphics[width=\linewidth]{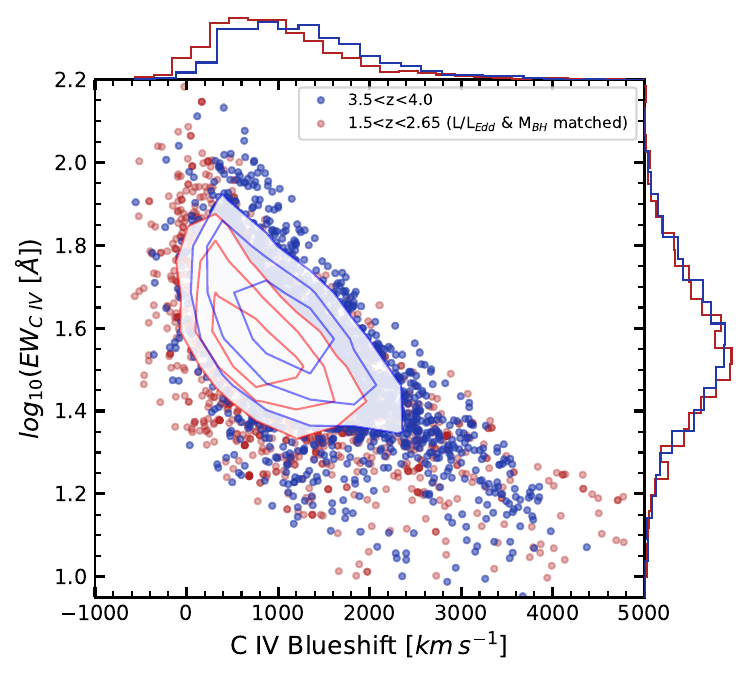}
 \caption{The \ion{C}{iv} emission space for the $3.5<z<4.0$ sample (blue) and the M$_{\textnormal{BH}}$ and L/L$_\textnormal{Edd}$ matched $1.5<z<2.65$ quasars (red) from \citet{2023MNRAS.523..646T}. Density contours encircle  68, 50 and 25 per cent of the sample, respectively.}
 \label{fig:CIV-MBH_LEDD_matched}
\end{figure}

We observe good agreement in the \ion{C}{iv} emission line properties when matching the $3.5<z<4.0$ sample to the $1.5<z<2.65$ quasars in L$_{3000}$ or both M$_{\textnormal{BH}}$ and L/L$_\textnormal{Edd}$. The distributions are illustrated in Fig. \ref{fig:CIV-L3000_matched} and Fig. \ref{fig:CIV-MBH_LEDD_matched} respectively. In both cases, the marginalised distributions in \ion{C}{iv} EW are in good agreement between $1.5<z<2.65$ and $3.5<z<4.0$. The high \ion{C}{iv} blueshift tail of the $3.5<z<4.0$ sample is much more consistent with the $1.5<z<2.65$ quasars when matching in either L$_{3000}$ or \emph{both} M$_{\textnormal{BH}}$ and L/L$_\textnormal{Edd}$, compared to the marginalised distributions presented in Fig. \ref{fig:CIV-Unmatched}. However, the number of objects with \ion{C}{iv} blueshift $\leq 500 \textnormal{km\,s}^{-1}$ tails off more rapidly at $3.5<z<4.0$. In Appendix \ref{Appendix:Zsys} we demonstrate that this can be attributed to the reduced rest-frame wavelength coverage of the $3.5<z<4.0$ spectra compared to the $1.5<z<2.65$ spectra for systemic redshift estimation. Truncating the $1.5<z<2.65$ spectra to the same rest-frame wavelength range as our $3.5<z<4.0$ spectra (namely $\lambda < 2000$\AA\,) results in a slight over-estimation of $z_{\textnormal{sys}}$ at lower \ion{C}{iv} blueshifts. At these rest-frame wavelengths, we no longer have access to the \ion{Mg}{ii}\,$\lambda2800$ emission line for calculation of $z_{\textnormal{sys}}$. $\ion{Mg}{ii}$ is not affected by narrow associated absorption, which can affect the \ion{C}{iv} line profile as discussed in detail in Appendix \ref{Appendix:Zsys}. This explains the discrepancies observed in the 1-D marginalised \ion{C}{iv} blueshift histograms presented in both Fig. \ref{fig:CIV-L3000_matched} and Fig. \ref{fig:CIV-MBH_LEDD_matched}. The high blueshift end is not affected by the systemic redshift bias due to the reduced rest-frame wavelength coverage of the high-redshift quasar spectra. 

While the \ion{C}{iv} emission line properties appear to be consistent across the entire redshift range investigated, it is important to note that the SMBH mass estimates at $1.5<z<2.65$ were derived from the \ion{Mg}{ii} emission line \citep{2023MNRAS.523..646T}. As discussed in Section \ref{SEC:PQP}, the SMBH masses for the $3.5<z<4.0$ sample are derived from the \ion{C}{iv} emission with an associated correction to the FWHM to account for the line asymmetry \citep{2017MNRAS.465.2120C}. This could potentially lead to biases in the results presented in Fig. \ref{fig:CIV-MBH_LEDD_matched}. 

\begin{figure}
 \includegraphics[width=\linewidth]{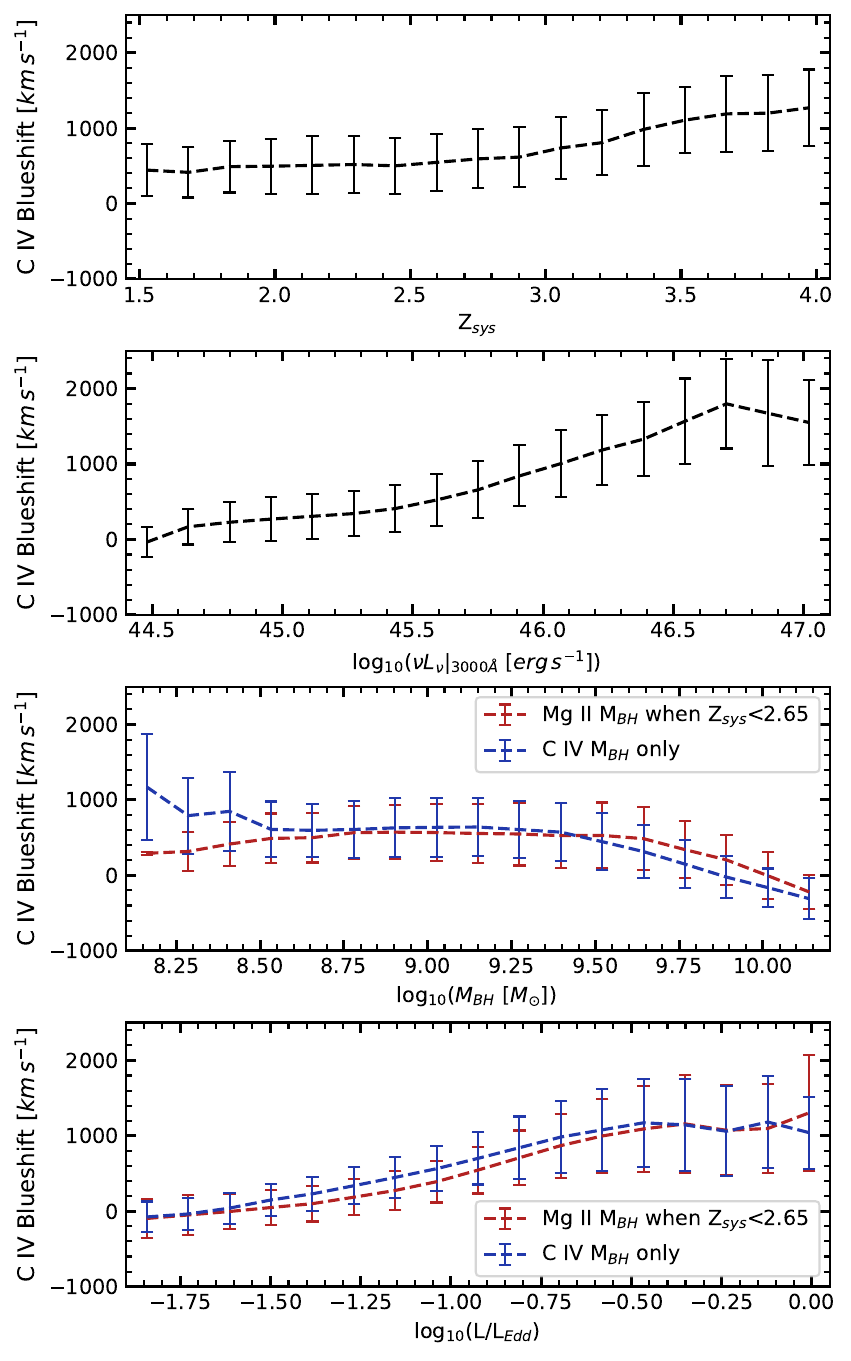} 
  \caption{The \ion{C}{iv} blueshift as a function of systemic redshift (top), UV continuum luminosity (top-middle), SMBH mass (bottom-middle) and Eddington-scaled accretion rate (bottom). The data is compressed into 17 equidistant bins, where we present the median and median absolute deviation (MAD) of each bin in the appropriate panels. We see a tendency for increasing \ion{C}{iv} blueshifts with increasing systemic redshift, UV continuum luminosity and Eddington-scaled accretion rate. The trends with UV continuum luminosity and Eddington-scaled accretion rate can explain why we observe an increase in \ion{C}{iv} blueshift as we tend to larger systemic redshifts. There is limited evidence of a strong trend between SMBH mass and \ion{C}{iv} blueshift (bottom middle), with a slight tendency to lower \ion{C}{iv} blueshifts as we approach the upper limit of our SMBH masses for the sample. There is limited evidence to suggest that the emission line used to estimate virial black hole masses will have any significant impact on our results.}
 \label{fig:Global_trends_CIV}
\end{figure}

In Fig. \ref{fig:Global_trends_CIV} we investigate how the \ion{C}{iv} blueshift evolves with redshift, luminosity, black hole mass and accretion rate. The \ion{C}{iv} blueshift increases with systemic redshift (top). However, when one accounts for the trends in blueshift with UV luminosity (top middle) and Eddington-scaled accretion rate (bottom), this trend can be explained by the higher luminosities probed at high redshifts due to the flux limits of the SDSS. Furthermore, when one considers the redshift evolution of the quasar luminosity function (QLF), whereby the number density of the brightest quasars exponentially decreases beyond $z\gtrsim3.5$ \citep[Fig.7;][]{2019MNRAS.488.1035K}, the increase in \ion{C}{iv} blueshift at $z\sim3-3.5$ (top) is consistent with the fact that we only see the most luminous quasars at these redshifts in a flux-limited sample. We also observe limited evolution in the \ion{C}{iv} blueshift as a function of SMBH mass (bottom middle) with a slight tendency for \ion{C}{iv} blueshifts to decrease with increasing SMBH masses. In addition, we observe no significant change to the trends in Fig. \ref{fig:Global_trends_CIV}, when we measure the black hole masses from the \ion{Mg}{ii} emission versus the \ion{C}{iv} emission line with a blueshift dependent correction applied to the \ion{C}{iv} FWHM (Section \ref{sec:Methods}; \citealt{2017MNRAS.465.2120C}). 

\subsection{\ion{He}{ii} emission}\label{sec:HeII}

\begin{figure}
 \includegraphics[width=\linewidth]{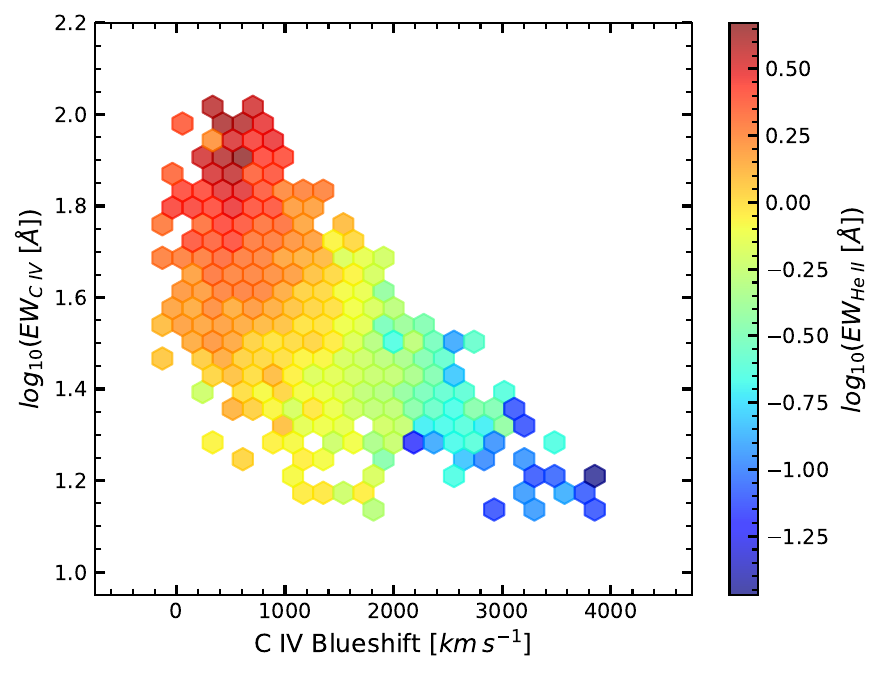}
 \caption{The median observed \ion{He}{ii} EW in bins of \ion{C}{iv} blueshift and \ion{C}{iv} EW for the $3.5<z<4.0$ sample. The \ion{He}{ii} EW is correlated with both \ion{C}{iv} blueshift and \ion{C}{iv} EW, with the most blueshifted \ion{C}{iv} lines only observed when the \ion{He}{ii} EW is low. Conversely, the highest EW \ion{C}{iv} lines are only observed when the \ion{He}{ii} EW is high.}
 \label{fig:CIV_HeII_hex}
\end{figure}

The \ion{He}{ii} line is generated by the direct recombination of \ion{He}{iii} ions to \ion{He}{ii} (with some covering factor dependence) and is therefore a "clean" measurement of the number of photons above 54.4 eV \citep[Section 2c;][]{1987ApJ...323..456M}. Hence, provided that the $\ion{He}{ii}$ line properties can be robustly measured, the $\ion{He}{ii}$ emission is a better indicator of the ionisation potential of the soft X-ray photons than \ion{C}{iv} \citep{2021MNRAS.504.5556T, 2023MNRAS.523..646T}. Through use of MFICA, we have been able to reconstruct the $\ion{He}{ii}$ emission line profile even at relatively modest signal-to-noise ratios (see Fig. \ref{fig:Spectra_SN} and Appendix \ref{Appendix:Shen}). \autoref{fig:CIV_HeII_hex} illustrates how the \ion{He}{ii} line properties change across the \ion{C}{iv} emission space. We recover the same trends as those uncovered by \citet{10.1093/mnras/stt2230} and \citet{2020MNRAS.492.4553R} at lower redshifts. \autoref{fig:CIV_HeII_hex} illustrates a clear anti-correlation between the \ion{He}{ii} EW and the \ion{C}{iv} blueshift as well as an additional correlation between the \ion{He}{ii} and \ion{C}{iv} EWs. We find that the same trends in \ion{He}{ii} EW and both \ion{C}{iv} EW and blueshift are recovered when one constructs composite stacked spectra, from 20 different regions in the \ion{C}{iv} emission space, and measures the line properties from the high signal-to-noise composites directly rather than using MFICA reconstructions. This confirms that despite the modest signal-to-noise of many of the individual spectra in the $3.5<z<4.0$ sample, the MFICA is still able to recover meaningful signal from the low EW emission lines.

\begin{figure}
 \includegraphics[width=\linewidth]{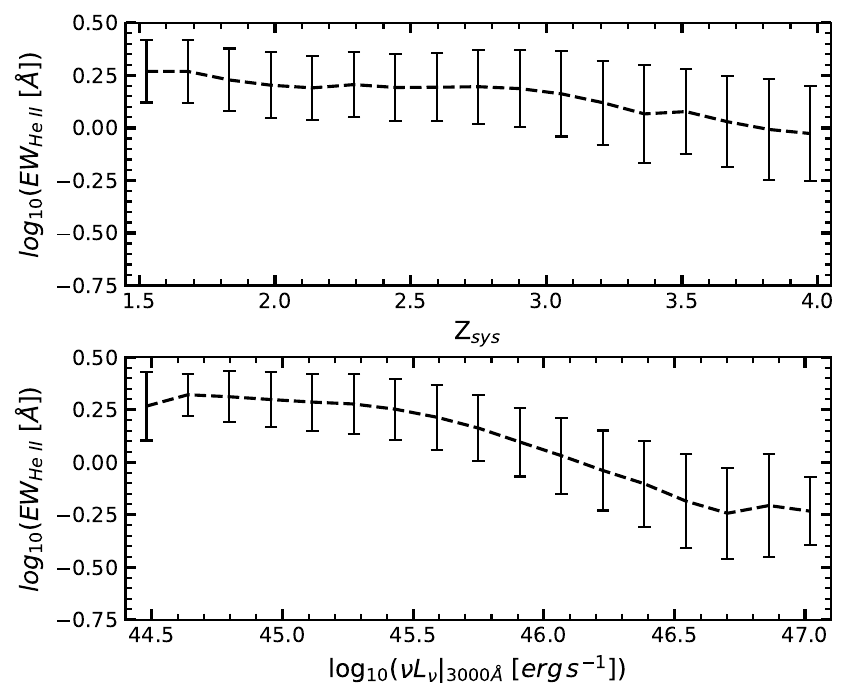} 
  \caption{The \ion{He}{ii} EW as a function of systemic redshift (top) and UV continuum luminosity (bottom). The data is compressed into 17 equidistant bins, where we present the median and median absolute deviation (MAD) of each bin in the appropriate panels. We see a slight tendency for decreasing \ion{He}{ii} EWs with increasing systemic redshift. The trends with \ion{He}{ii} EW and UV continuum luminosity can therefore well explain the apparent evolution with systemic redshift.}
 \label{fig:Global_trends_HeII}
\end{figure}

Under the assumption that the \ion{He}{ii} EW is a good proxy for the ionising flux at 54.4 eV (see Section \ref{sec:Outflow drivers} for full discussion), this result is consistent with the hypothesis that UV line-driven outflows give rise to the blueshift observed in the \ion{C}{iv} emission of quasars. A soft SED is a prerequisite for the strong blueshifting observed in the \ion{C}{iv} emission and over-ionisation of the outflowing material leads to weaker \ion{C}{iv} blueshifts. In Fig. \ref{fig:Global_trends_HeII} we present the evolution of the \ion{He}{ii} EW as a function of both systemic redshift and UV continuum luminosity. We observe no convincing evidence of an evolution in the \ion{He}{ii} EW with systemic redshift, save for a slight tendency to lower \ion{He}{ii} EW at $3.5<z<4.0$. However, this tendency is well accounted for when one considers the trend in \ion{He}{ii} EW with UV continuum luminosity, whereby the \ion{He}{ii} EW decreases with luminosity and therefore with systemic redshift as a result of the sampling bias at $z\gtrsim3.5$. Since the relation presented in Fig. \ref{fig:CIV_HeII_hex} is observed in quasars across the entire redshift range, $1.5<z<4.0$, and the \ion{He}{ii} EW demonstrates little to no evolution with systemic redshift in Fig. \ref{fig:Global_trends_HeII}, we can conclude that the SED properties of quasars remain consistent across the entire $1.5<z<4.0$ redshift range. 

\section{Discussion} \label{sec:discussion}
\subsection{Probing the driver of outflows} \label{sec:Outflow drivers}

We have explored trends in the rest-frame UV emission line properties of quasars at $1.5<z<4.0$, analysed using the same methodology across the full redshift range. As shown in Fig. \ref{fig:CIV-L3000_matched} and Fig. \ref{fig:CIV-MBH_LEDD_matched}, we find that the \ion{C}{iv} emission line properties, and in particular the high-blueshift tail at $3.5<z<4.0$ and $1.5<z<2.65$ are best matched when matching the quasar samples in either UV luminosity or equivalently \textit{both} SMBH mass and Eddington-scaled accretion rate. In Fig. \ref{fig:CIV_HeII_hex} we demonstrated that there is a strong and systematic trend between the \ion{C}{iv }blueshift and the \ion{He}{ii} EW in $3.5<z<4.0$ quasars, which is qualitatively very similar to the trend found in the $1.5<z<3.5$ population by e.g. \citet{2020MNRAS.492.4553R}. The result is broadly consistent with a paradigm where the \ion{C}{iv} blueshift is tracing a radiation line-driven wind with the ability to launch a wind anti-correlating with the number of high-energy ionizing photons above 54eV \citep{2015MNRAS.449.1593B}. We will therefore proceed from hereon under the assumption that the \ion{C}{iv} emission line properties encode information about the strength of a radiation line-driven wind. A key question of interest is how the wind properties then relate to fundamental properties of the quasar such as UV luminosity, SMBH mass and Eddington-scaled accretion rate. Recently,  \citet{2023MNRAS.523..646T} have looked at this exact question using a sample of $\sim$190,000 SDSS quasars at $1.5<z<2.65$. We can therefore extend the \citet{2023MNRAS.523..646T} analysis to explicitly ask if the same fundamental properties drive trends in quasar UV emission line properties at $3.5<z<4.0$.

\begin{figure}
 \includegraphics[width = \linewidth]{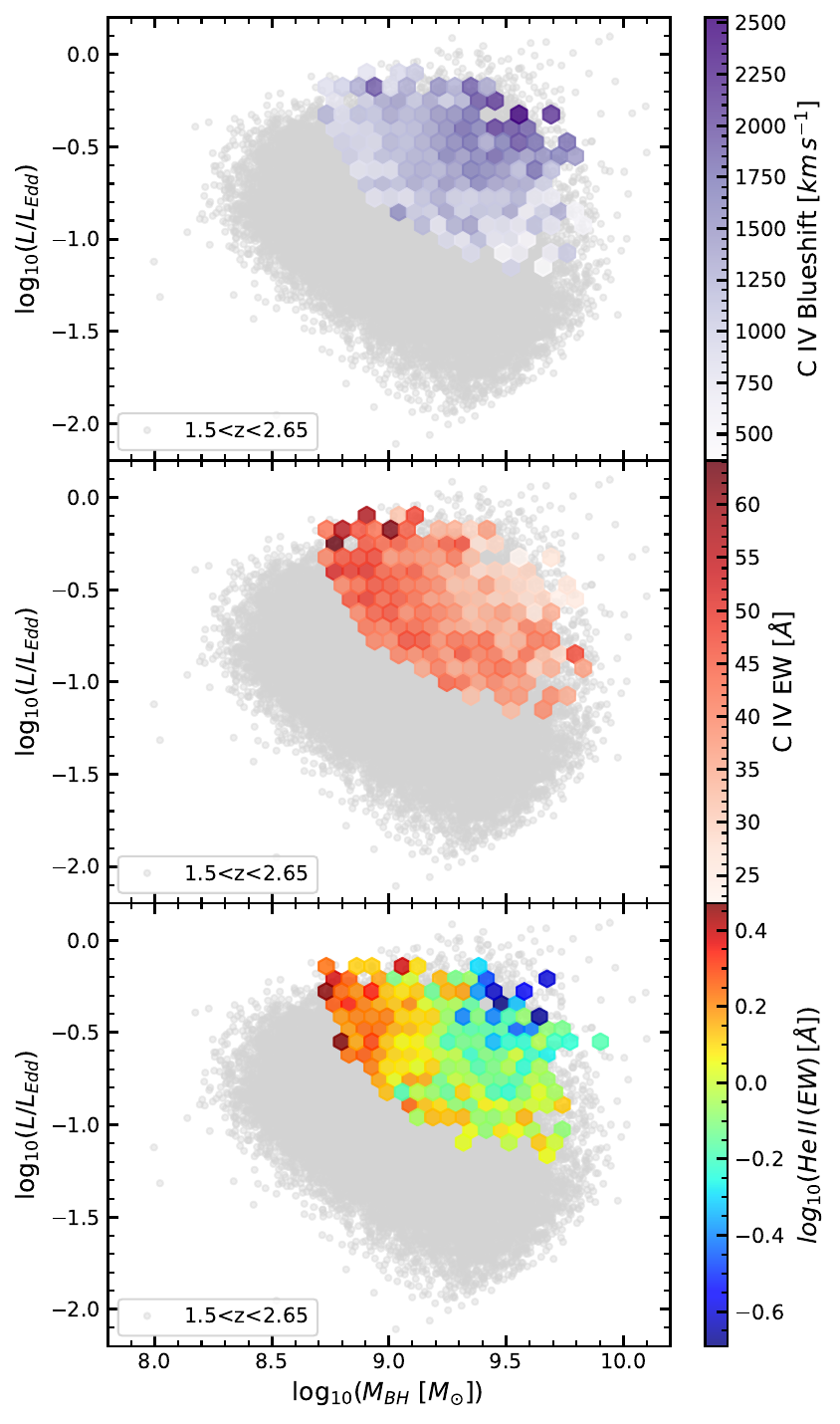}
 \caption{The median observed \ion{C}{iv} blueshift (top), \ion{C}{iv} EW (middle) and \ion{He}{ii} EW (bottom) in bins of SMBH mass and Eddington-scaled accretion rate for the $3.5<z<4.0$ sample overlaid on the M$_\textsc{bh}$ - L/L$_\textnormal{Edd}$ plane at $1.5<z<2.65$ \citep[grey,][]{2023MNRAS.523..646T}. Save for a clear flux limit, the $1.5<z<2.65$ and $3.5<z<4.0$ samples are well-matched in this space. We note that above L/L$_\textnormal{Edd} \sim$ 0.2 there is evidence of a clear trend between \ion{He}{ii} EW and SMBH mass. Below $\sim 10^9 M_\odot$, the quasars generally have stronger UV emission lines and weaker \ion{C}{iv} blueshifts and above this threshold, the contrary is true.}
 \label{fig:Mass_Accretion_space}
\end{figure}

\autoref{fig:Mass_Accretion_space} illustrates the M\textsc{bh} - L/L$_\textnormal{Edd}$ plane, for both the $1.5<z<2.65$ and $3.5<z<4.0$ samples, with trends in \ion{C}{iv} blueshift, \ion{C}{iv} EW and $\ion{He}{ii}$ EW at $3.5<z<4.0$ overlaid. As expected, we observe a clustering at $3.5<z<4.0$ in the top right-hand corner of the parameter space and the $3.5<z<4.0$ sample is biased to higher black hole masses and Eddington-scaled accretion rates compared to $1.5<z<2.65$, due to the flux limits of the SDSS survey. Crucially though, and in contrast to quasars at even higher redshifts as we discuss later, the $3.5<z<4.0$ sample overlaps considerably with the $1.5<z<2.65$ sample in SMBH mass and Eddington-scaled accretion rate.

\begin{figure}
 \includegraphics[width =\linewidth]{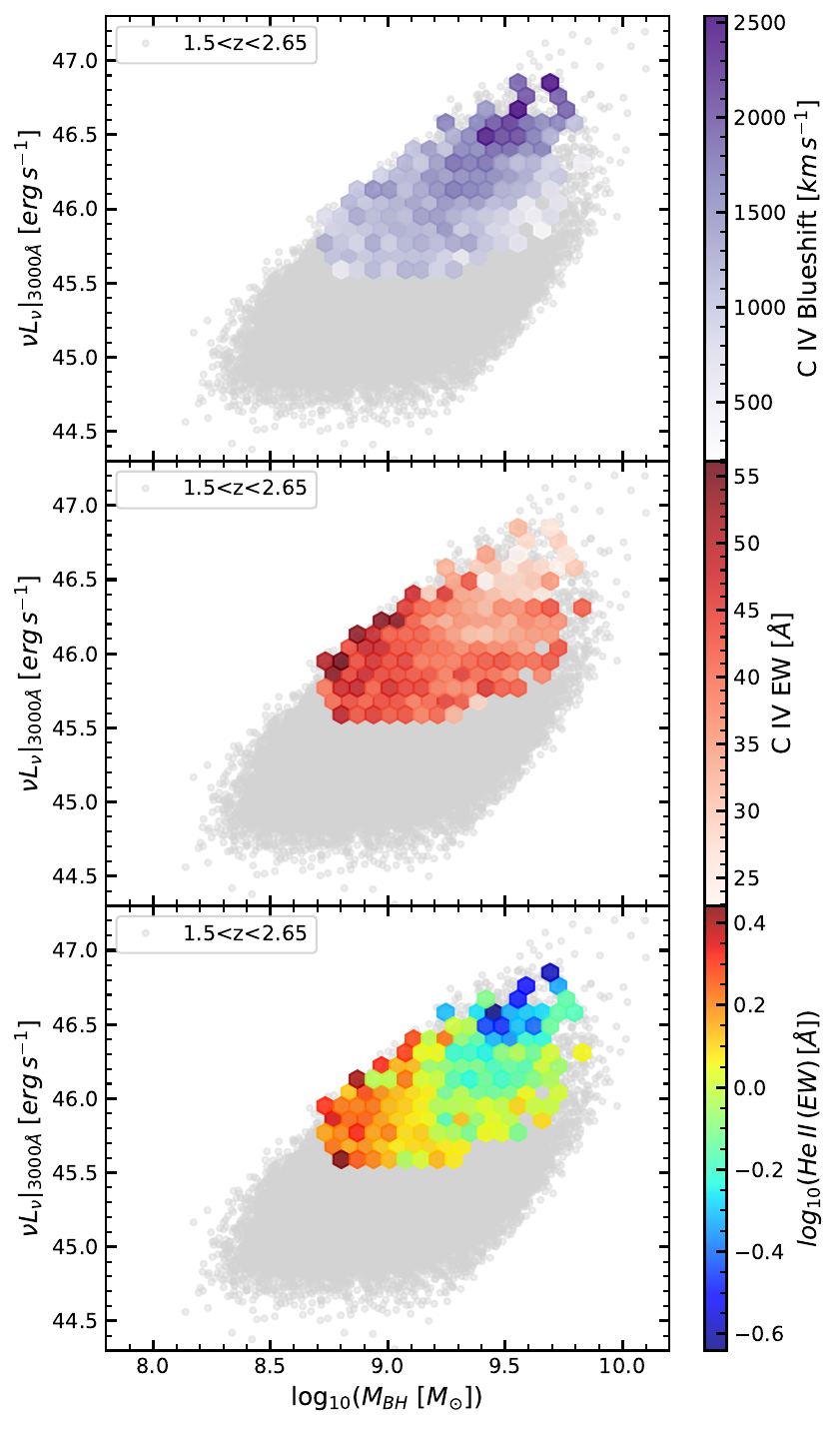} 
 \caption{The median observed \ion{C}{iv} blueshift (top), \ion{C}{iv} EW (middle) and \ion{He}{ii} EW (bottom) in bins of SMBH mass and 3000\AA\, UV continuum luminosity for the $3.5<z<4.0$ sample overlaid on the M$_\textsc{bh}$ - L$_{3000}$ plane at $1.5<z<2.65$ \citep[grey,][]{2023MNRAS.523..646T}. A clear flux limit is observed at $\nu L\nu|_{3000\AA} \sim 45.5\,\textnormal{erg\,s}^{-1}$ in the $3.5<z<4.0$ sample. Apart from that the $1.5<z<2.65$ and $3.5<z<4.0$ samples are well-matched in this space. Below $\sim 10^9 M_\odot$ the quasars have stronger UV emission lines and smaller \ion{C}{iv} blueshifts. Above this threshold, the contrary is true.}
 \label{fig:Mass_L3000_space}
\end{figure}

As is the case in the $1.5<z<2.65$ sample, we observe the strongest outflows above L/L$_\textnormal{Edd} \geq$ 0.2 and M\textsc{bh} $\geq 10^9 M_\odot$. Indeed, the trends in \ion{C}{iv} blueshift, \ion{C}{iv} EW and \ion{He}{ii} EW are consistent with the $1.5<z<2.65$ sample results, presented in \citet{2023MNRAS.523..646T}, suggesting that the underlying drivers of these UV line properties do not evolve with redshift. We also investigate trends in the M$_\textsc{bh}$ - L$_{3000}$ plane in Fig. \ref{fig:Mass_L3000_space}. The flux limit of the $3.5<z<4.0$ sample is evident at $\nu L\nu|_{3000\AA} \sim 45.5\,\textnormal{erg\,s}^{-1}$, the threshold above which quasar feedback is considered to be effective \citep{2014MNRAS.442..784Z}. The trends in line properties in the M$_\textsc{bh}$ - L$_{3000}$ plane are also consistent with those seen in $1.5<z<2.65$ quasars by \citet{2023MNRAS.523..646T}. A key result of our work is therefore explicitly demonstrating that redshift is not a fundamental parameter in determining quasar UV emission line (and by implication outflow) properties and that these properties are instead governed by SMBH mass and accretion rate. Moreover, the results are not sensitive to the emission line used for the SMBH mass estimates, which demonstrates that with large enough statistical samples global trends of UV line properties with SMBH mass and accretion rate can be recovered even when one uses \ion{C}{iv} lines for SMBH mass estimates (e.g, Fig. \ref{fig:Global_trends_CIV}). 

\citet{2019A&A...630A..94G} have proposed that accreting black holes with a mass, M\textsc{bh} $\geq 10^8 M_\odot$, and an Eddington-scaled accretion rate, L/L$_\textnormal{Edd} \geq 0.25$ are expected to facilitate strong radiation-driven winds. Conversely, objects whose Eddington-scaled accretion rate falls below this threshold will likely produce failed line-driven disc winds, or in extreme cases, the outflowing material is magnetically-driven, and the feedback kinetic. \citet{2023MNRAS.523..646T} find that the observations at $1.5<z<2.65$ are in good agreement with the proposed framework, only measuring significant \ion{C}{iv} blueshifts ($\geq 1000\,\textnormal{km\,s}^{-1}$), at $1.5<z<2.65$, when objects are both strongly-accreting and high mass. We confirm that at $3.5<z<4.0$, the \ion{C}{iv} blueshift is again strongly dependent on the mass and accretion rate in much the same way as for the $1.5<z<2.65$ population and that at high accretion rates and lower SMBH masses, the ionising potential above 54.4 eV, as traced by the \ion{He}{ii} EW, prevents strong disc winds from being launched.

Fundamentally, the similarity in the trends in UV emission line properties with luminosity, mass and accretion rate between $1.5<z<2.65$ and $3.5<z<4.0$, strongly suggests that the same accretion and wind-driving mechanisms are at play across the entire $1.5<z<4.0$ redshift range. We will now discuss the implication of these results for studies of quasar winds and outflows at even higher redshifts, where there have been claims that the UV emission line properties do indeed show some evolution \citep{2019MNRAS.487.3305M, Schindler_2020,2021ApJ...923..262Y, 2022ApJ...941..106F}.

\subsection{Implications at higher redshifts}

\begin{figure*} 
 \includegraphics[width=\linewidth]{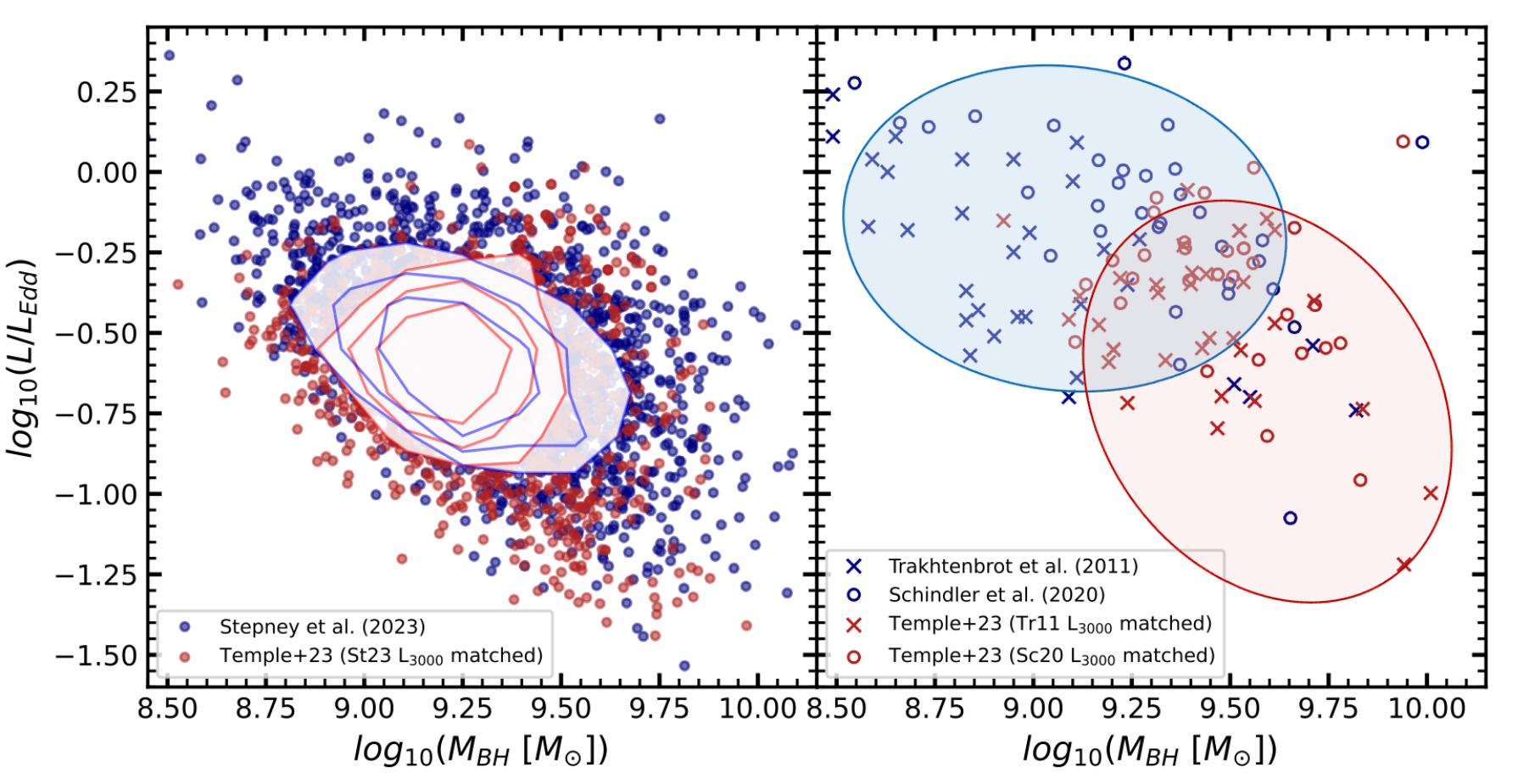}
 \caption{The M$_\textsc{bh}$ - L/L$_\textnormal{Edd}$ plane for the $3.5<z<4.0$ and corresponding UV luminosity matched $1.5<z<2.65$ samples (left) and the z$\sim$6 and corresponding UV luminosity matched $1.5<z<2.65$ samples (right). Density contours (left) encircle  68, 50 and 25 per cent of the sample, respectively. The cartoon ellipses (right) approximate the locations of the $z\sim6$ quasars (blue) and the $1.5<z<2.65$ quasars, when UV luminosity matched to the $z\sim6$ sample (red). When UV luminosity matching the $3.5<z<4.0$ sample to the $1.5<z<2.65$ quasars, the samples are consistent in the M$_\textsc{bh}$ - L/L$_\textnormal{Edd}$ plane, explaining the results presented in Fig. \ref{fig:CIV-L3000_matched} and Fig. \ref{fig:CIV-MBH_LEDD_matched}. When UV luminosity matching the z$\sim$6 sample to the $1.5<z<2.65$ quasars, we see an inconsistency between the samples in the M$_\textsc{bh}$ - L/L$_\textnormal{Edd}$ plane, with the $z\sim6$ objects biased towards higher L/L$_\textnormal{Edd}$ and lower M$_\textsc{bh}$ than their $1.5<z<2.65$ counterparts.}
 \label{fig:Temple_space}
\end{figure*}

Previous studies of quasars at the highest redshifts of $z\gtrsim6$ have concluded that their emission line properties, and in particular \ion{C}{iv} blueshifts are on average higher than quasars at lower redshifts even when the samples are matched in UV luminosity. We now consider quasar samples at $z\sim4.8$ from \citet{2011ApJ...730....7T} and at $z\gtrsim6$ from \citet{Schindler_2020} for explicit comparison to our work. As has been done in the literature, we construct a control sample of $1.5<z<2.65$ quasars drawing from the sample in \citet{2023MNRAS.523..646T}, and matching to the high-redshift samples in UV luminosity. In Fig. \ref{fig:CIV-L3000_matched} we found that matching the $3.5<z<4.0$ and $1.5<z<2.65$ samples in UV-luminosity resulted in a good agreement in the \ion{C}{iv} emission properties. The left-hand panel of Fig. \ref{fig:Temple_space} illustrates that the UV luminosity matching across the two redshift bins, results in a broadly consistent mapping to the M$_\textsc{bh}$ - L/L$_\textnormal{Edd}$ plane, which we have already determined in Section \ref{sec:Outflow drivers} are the fundamental parameters driving trends in \ion{C}{iv} blueshift. However, when we consider the \citet{2011ApJ...730....7T} and \citet{Schindler_2020} quasars at $z\sim5-7$, the UV-luminosity matched control samples no longer map onto the same region of the M$_\textsc{bh}$ - L/L$_\textnormal{Edd}$ plane as the high-redshift quasars, as illustrated in the right-hand panel of Fig. \ref{fig:Temple_space}. This is perhaps unsurprising when one considers the degeneracy in matching quasar samples through their luminosities, i.e, samples with low Eddington-scaled accretion rates and high SMBH masses will populate the same region of the luminosity distribution as samples with high Eddington-scaled accretion rates and low SMBH masses. As there are relatively small numbers of $z\sim5-7$ quasars and they are observed at higher Eddington-scaled accretion rates and lower SMBH masses, a region that isn't populated well by the $1.5<z<2.65$ sample used as a control, UV-luminosity matching is ineffective in selecting a control sample of quasars with a similar SMBH mass and Eddington-scaled accretion rate distribution as the quasars found at $1.5<z<2.65$. Indeed, \citet{2022ApJ...941..106F} do find that the $z>6$ quasars have systematically higher Eddington-scaled accretion rates compared to a UV-luminosity matched control sample at lower redshifts. Regardless of whether this is an intrinsic property of the highest redshift quasar population or a selection bias in the observed samples, this could in part explain the observed discrepancies in the \ion{C}{iv} blueshifts between $1.5<z<2.65$ and $z\gtrsim6$ quoted in the literature.

It is curious then, when exploring the trends in Fig. \ref{fig:Mass_Accretion_space}, that an extrapolation of these trends would imply that the lower mass, higher Eddington-scaled accretion rate quasars at $z>6$ should have higher \ion{He}{ii} EWs and therefore smaller \ion{C}{iv} blueshifts compared to the luminosity-matched control at $1.5<z<2.65$, contrary to what is observed. Unfortunately the \ion{He}{ii} EWs have not been measured for the highest redshift quasars, which precludes any direct comparison to the $3.5<z<4.0$ sample analysed here. In the case of the $3.5<z<4.0$ sample, our use of the novel MFICA technique has enabled us to extract information about the \ion{He}{ii} EWs even for spectra of modest signal-to-noise. While we do not have information on the \ion{He}{ii} properties at the highest redshifts, several papers have looked at the X-ray properties of the highest redshift quasars and confirmed the well-known correlation between $\alpha_{\textsc{ox}}$, the UV-to-Xray SED slope, and the UV luminosity. Higher luminosity quasars have proportionally lower X-ray flux and therefore softer ionising SEDs with no apparent evolution in $\alpha_{\textsc{ox}}$ with redshift \citep{2017A&A...603A.128N, 2020MNRAS.491.3884P}. 

If the trends illustrated in Fig. \ref{fig:Mass_Accretion_space} do break down for the $z\sim6$ population, this might point to a change in the underlying mechanisms driving the \ion{C}{iv} blueshift. \citet{2019MNRAS.487.3305M} have speculated about the role of orientation and obscuration in explaining the apparent lack of $z>6$ quasars with modest blueshifts. In a flux-limited sample probing only the highest UV-luminosity sources, quasars that are viewed more edge-on and/or more obscured would preferentially drop out of the sample. If the wind geometry is polar, these missing quasars would also be those at more modest blueshifts, thus explaining the bias to high blueshifts at $z>6$. Our work now explicitly demonstrates that despite sampling higher UV-luminosity sources at $3.5<z<4.0$, there is no such bias in their \ion{C}{iv} emission line properties relative to $z\sim2$, which could imply that the geometry and obscured fraction of quasars dramatically changes from $z\sim4$ to $z\gtrsim6$. Alternatively, it might be inappropriate to extrapolate the trends in Fig. \ref{fig:Mass_Accretion_space} to the super-Eddington regime given the inner accretion disc transitions from geometrically thin and optically thick to a slim disc as radiation pressure causes the disc to ``puff up" \citep{2019A&A...630A..94G}.   

It is also important to highlight that when illustrating trends in Fig. \ref{fig:Temple_space}, all the SMBH masses, luminosities and Eddington-scaled accretion rates for the \citet{2011ApJ...730....7T} sample were re-derived using Eqns \ref{eq:CIV_Correction} and \ref{eq:CIV_MASS}. Although we currently take the equivalent properties for the $z\gtrsim6$ quasars directly from \citet{Schindler_2020}, our results would remain unchanged if we used the continuum luminosities and FWHM of the emission lines from that paper to re-derive black hole masses and Eddington-scaled accretion rates for the $z>6$ sample. Considerable care has also been taken in this work to ensure that the $3.5<z<4.0$ and $1.5<z<2.65$ samples have employed the same methodologies for calculating quasar systemic redshifts and measuring UV line properties. This was beyond the scope of the current paper for the quasar samples analysed at higher redshifts and might result in biases when comparing to the lower redshift population (see e.g. Appendix \ref{Appendix:Shen}).

A key conclusion of our study is that, under the premise that the \ion{C}{iv} emission line blueshift is tracing the velocity of radiatively driven disc winds, the fundamental parameters that govern the blueshift are the SMBH mass and Eddington-scaled accretion rate. We have shown that quasars at the highest redshifts do not map on to the SMBH mass Eddington-scaled accretion rate plane occupied by similarly luminous quasars at lower redshifts. A more complete sampling of this plane is clearly needed at $z>6$ to determine whether the observed evolution in UV-line properties is due to an intrinsic difference in the masses and accretion rates of the first quasars, or conversely, if different mechanisms are at play in driving the observed \ion{C}{iv} blueshift compared to quasars at $1.5<z<4.0$. This is now becoming possible with the launch of the \textit{JWST} which will enable lower mass, lower accretion rate quasars to be found in the very early Universe, as well as robustly measuring black hole masses and accretion rates for known $z>6$ quasars using Balmer lines, which suffer considerably less from systematic biases.

\section{Conclusions} 

We have analysed the rest-frame UV spectra of 2,531 $3.5<z<4.0$ quasars from the SDSS DR16Q catalogue and studied the evolution of the rest-frame UV properties of quasars in the redshift range $1.5<z<4.0$. 

\begin{itemize}
    \item {We used high signal-to-noise template spectra of quasars at $1.5<z<3.5$ and a cross-correlation algorithm to calculate updated systemic redshifts for the $3.5<z<4.0$ quasars. The templates take into account the known systematic velocity offsets between different emission lines as a function of quasar properties as well as the diversity of the \ion{C}{iv} emission line morphologies. This enables accurate systemic redshift estimates using just the rest-frame UV, a technique that can now be applied to quasars at even higher redshifts}. 

    \item{We use Mean Field Independent Component Analysis (MFICA) to produce high signal-to-noise reconstructions of the individual quasar spectra from which we measure non-parametric emission line properties. We recover the same trends in \ion{C}{iv} EW and \ion{C}{iv} blueshift as those reported at $1.5<z<2.65$. We find that there is no evidence for evolution in the \ion{C}{iv} blueshifts and EWs between $1.5<z<2.65$ and $3.5<z<4.0$ when matching the quasars in either UV continuum luminosity, L$_{3000}$, or \emph{both} SMBH mass and Eddington-scaled accretion rate.}

    \item{The use of MFICA enables us to reconstruct the \ion{He}{ii} emission line profile even in modest signal-to-noise spectra. We recover the well-known correlation between the \ion{He}{ii} and \ion{C}{iv} EWs as well as an anti-correlation between the \ion{He}{ii} EW and \ion{C}{iv} blueshift. Under the assumption that the \ion{C}{iv} blueshift traces broad line region outflows, we conclude that the quasar SED and more specifically the ionising flux above 54eV is a key determinant of the ability to launch outflows. Moreover, there is a common SED-dependent mechanism for quasar-driven outflows at play in quasars over the entire redshift range $1.5<z<4.0$.}

    \item{We examine how the rest-frame UV line properties depend on fundamental properties of the quasars - namely L$_{3000}$, M$_\textsc{bh}$ and L/L$_\textnormal{Edd}$. The $3.5<z<4.0$ quasars are more luminous than the $1.5<z<2.65$ quasars but show the same trends in their emission line properties with SMBH mass and Eddington-accretion rate as their lower redshift counterparts. As reported by \citet{2023MNRAS.523..646T}, significant \ion{C}{iv} blueshift measurements require \emph{both} L/L$_\textnormal{Edd} \geq$ 0.2 and M\textsc{bh} $\geq 10^9 M_\odot$. Likewise, when L/L$_\textnormal{Edd} \geq$ 0.2, we observe a clear evolution in the \ion{He}{ii} EW, with higher SMBH mass objects presenting both weaker \ion{He}{ii} emission and larger \ion{C}{iv} blueshifts.} 

    \item{We explicitly show that matching quasars at $3.5<z<4.0$ to those at $1.5<z<2.65$ based on their UV continuum luminosity, ensures a consistent mapping of quasars in both redshift bins on to the M$_\textsc{bh}$ - L/L$_\textnormal{Edd}$ plane. However, when considering the much smaller sample of quasars at $z\sim5-7$ where UV-emission line properties have been measured, the UV-luminosity matched $1.5<z<2.65$ samples are biased to higher SMBH mass and lower Eddington-scaled accretion rates than observed at the highest redshifts. If SMBH mass and Eddington-scaled accretion rate are indeed the fundamental parameters driving the \ion{C}{iv} blueshift, this might at least partially explain the observed evolution of \ion{C}{iv} blueshift seen in the highest redshift quasars. We therefore conclude that matching quasar samples in different redshift bins using the UV continuum luminosity is only viable when one can also achieve a reasonable match in \emph{both} the SMBH mass and Eddington-scaled accretion rate.}

    \item{We hypothesise a number of explanations for the inconsistent distributions in the M$_\textsc{bh}$ - L/L$_\textnormal{Edd}$ plane between $1.5<z<4.0$ and $z\sim5-7$. \citet{2022ApJ...941..106F} find that quasars at $z\gtrsim6$ have a tendency towards higher Eddington-scaled accretion rates and lower SMBH masses than their $1.5<z<2.65$ UV luminosity-matched counterparts. Interestingly, if the relationship between the \ion{He}{ii} EW and the M$_\textsc{bh}$ - L/L$_\textnormal{Edd}$ plane persists to $z\sim5-7$, we would expect objects with higher Eddington-scaled accretion rates and lower SMBH masses to exhibit weaker \ion{C}{iv} blueshifts. This trend is contrary to the observations at $z\gtrsim6$, for which we suggest two possible explanations. One possibility is that the relationships between the M$_\textsc{bh}$, L/L$_\textnormal{Edd}$ and the UV emission line properties, at $1.5<z<4.0$, do not hold at higher redshifts where the Eddington ratios appear higher, and hence the mechanisms driving outflows at $z\gtrsim6$ are intrinsically different. Alternatively, the different line-fitting methodologies used to analyse the quasar samples at $z\gtrsim6$ may prevent a direct comparison to the UV emission line properties at more modest redshifts.}
\end{itemize}

\section*{Acknowledgements}

We thank Prof. Christian Knigge and the anonymous referee for their useful comments that helped strengthen the work. 

MS acknowledges funding from the University of Southampton via the Mayflower studentship. MB acknowledges funding from the Royal Society via a University Research Fellowship (UF160074). MJT acknowledges support from a FONDECYT postdoctoral fellowship (3220516). ALR acknowledges support from UKRI (MR/T020989/1). For the purpose of open access, the author has applied a Creative Commons Attribution (CC BY) licence to any Author Accepted Manuscript version arising from this submission.

Funding for the Sloan Digital Sky Survey IV has been provided by the Alfred P. Sloan Foundation, the U.S. Department of Energy Office of Science, and the Participating Institutions. SDSS-IV acknowledges support and resources from the Center for High-Performance Computing at the University of Utah. The SDSS web site is www.sdss.org\footnote{\url{www.sdss.org}}.

SDSS-IV is managed by the Astrophysical Research Consortium for the 
Participating Institutions of the SDSS Collaboration including the 
Brazilian Participation Group, the Carnegie Institution for Science, 
Carnegie Mellon University, the Chilean Participation Group, the French Participation Group, Harvard-Smithsonian Center for Astrophysics, 
Instituto de Astrof\'isica de Canarias, The Johns Hopkins University, Kavli Institute for the Physics and Mathematics of the Universe (IPMU) / 
University of Tokyo, the Korean Participation Group, Lawrence Berkeley National Laboratory, 
Leibniz Institut f\"ur Astrophysik Potsdam (AIP),  
Max-Planck-Institut f\"ur Astronomie (MPIA Heidelberg), 
Max-Planck-Institut f\"ur Astrophysik (MPA Garching), 
Max-Planck-Institut f\"ur Extraterrestrische Physik (MPE), 
National Astronomical Observatories of China, New Mexico State University, 
New York University, University of Notre Dame, 
Observat\'ario Nacional / MCTI, The Ohio State University, 
Pennsylvania State University, Shanghai Astronomical Observatory, 
United Kingdom Participation Group,
Universidad Nacional Aut\'onoma de M\'exico, University of Arizona, 
University of Colorado Boulder, University of Oxford, University of Portsmouth, 
University of Utah, University of Virginia, University of Washington, University of Wisconsin, 
Vanderbilt University, and Yale University.

%%%%%%%%%%%%%%%%%%%%%%%%%%%%%%%%%%%%%%%%%%%%%%%%%%
\section*{Data Availability}

The spectroscopic data underlying this article are available from SDSS\footnote{\url{https://www.sdss4.org/dr17/}}.

%%%%%%%%%%%%%%%%%%%% REFERENCES %%%%%%%%%%%%%%%%%%

\bibliographystyle{mnras}
\bibliography{HighZ_Quasars} 

%%%%%%%%%%%%%%%%% APPENDICES %%%%%%%%%%%%%%%%%%%%%

\appendix 

\section{The MFICA Components}\label{Appendix:MFICA}

\begin{figure*}
 \includegraphics[width=\linewidth]{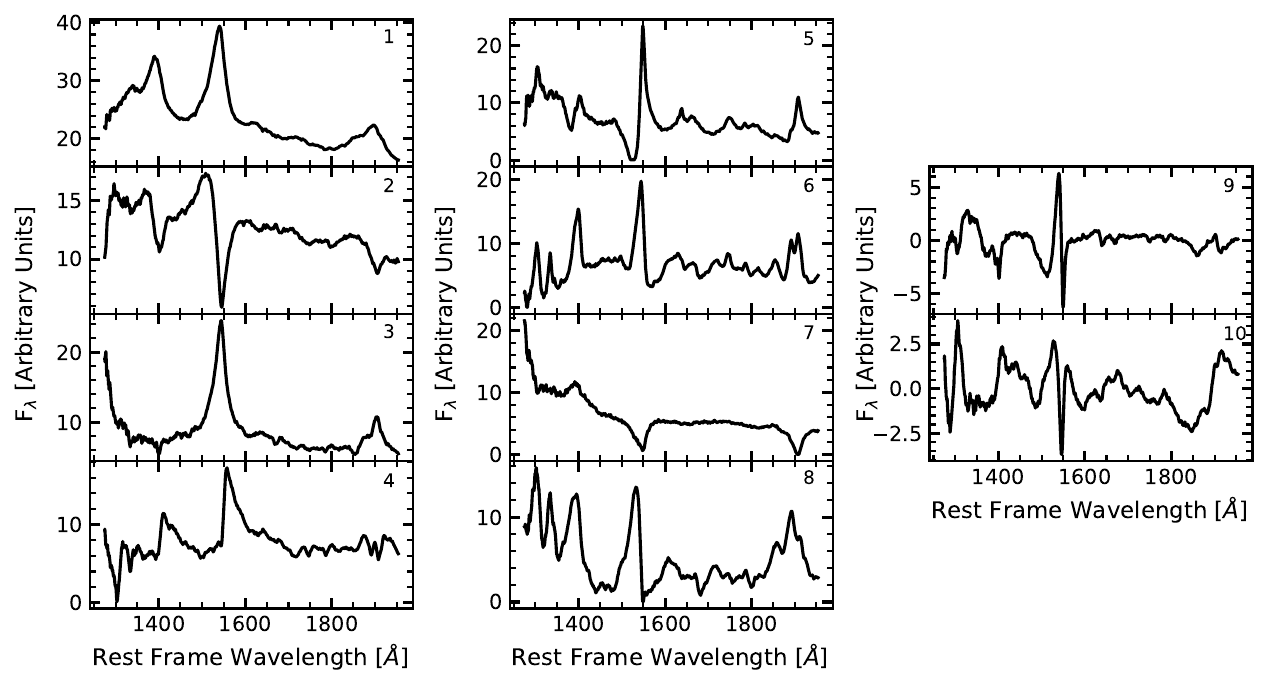} 
 
  \caption{The MFICA components, generated using the technique described in \citet{2013MNRAS.430.3510A}, used to reconstruct quasar spectra whose \ion{C}{iv} EW $<20$\AA\,. Components "9" and "10" are correction components and are therefore permitted to have negative weights in the MCMC fitting.}
 \label{fig:COMP_low}
\end{figure*}

\begin{figure*}
 \includegraphics[width=\linewidth]{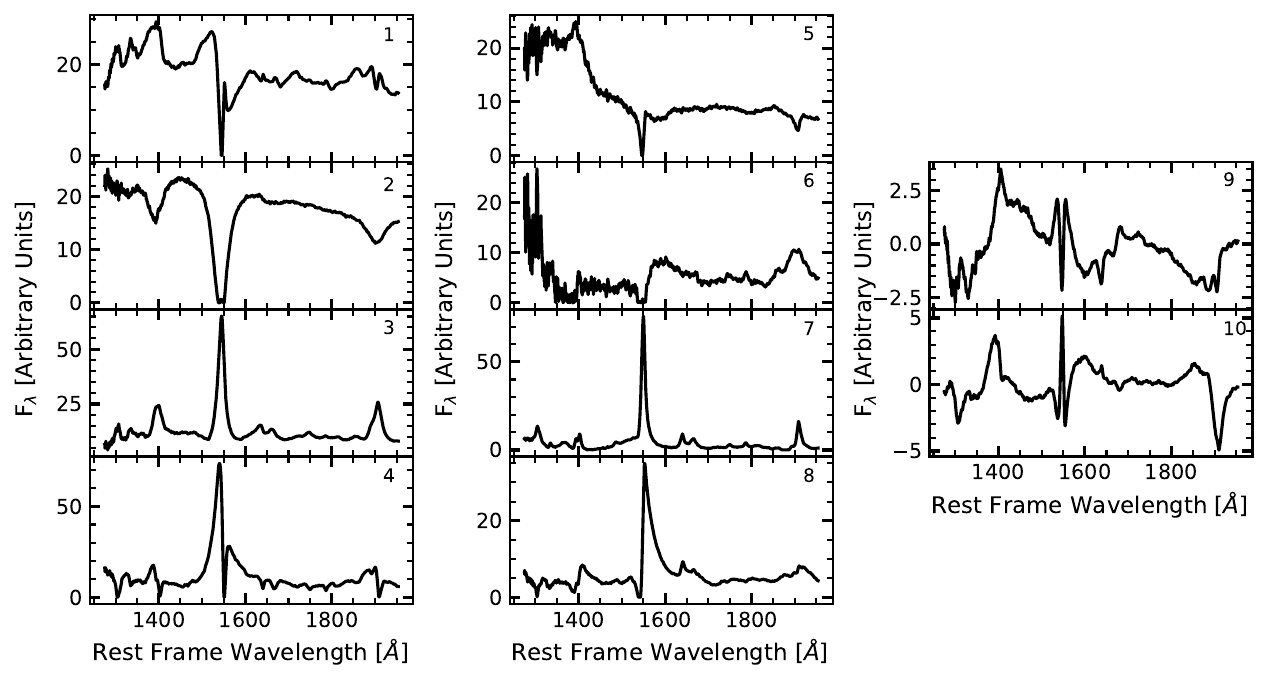} 
 
  \caption{The MFICA components, generated using the technique described in \citet{2013MNRAS.430.3510A}, used to reconstruct quasar spectra whose $20<\,$\ion{C}{iv} EW$<70$\AA\,. Components "9" and "10" are correction components and are therefore permitted to have negative weights in the MCMC fitting.}
 \label{fig:COMP_int}
\end{figure*}

\begin{figure*}
 \includegraphics[width=\linewidth]{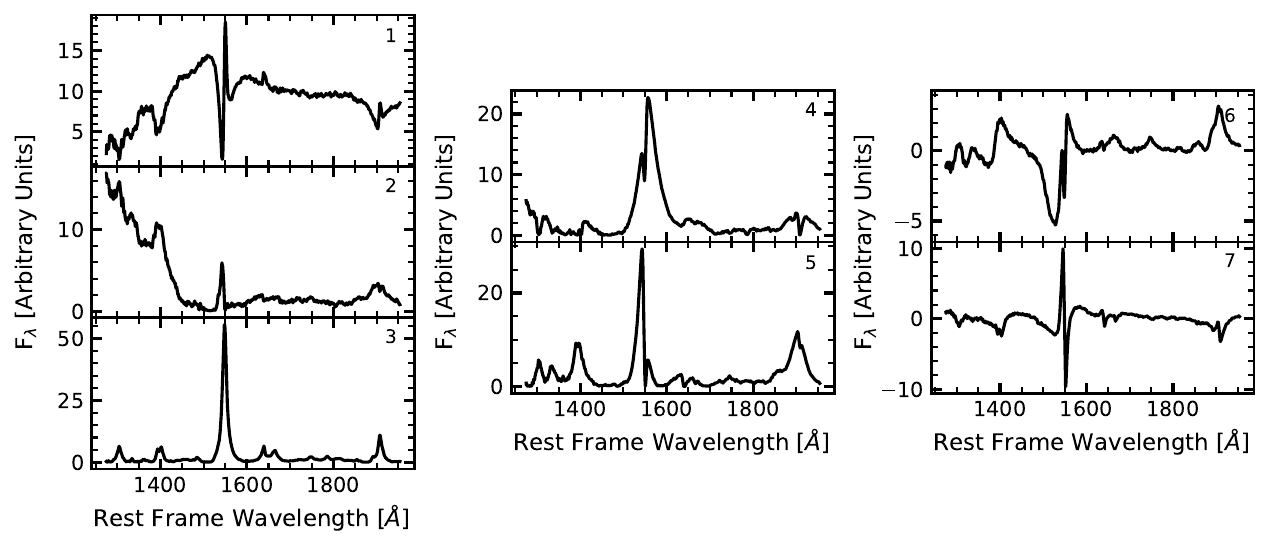} 
 
  \caption{The MFICA components, generated using the technique described in \citet{2013MNRAS.430.3510A}, used to reconstruct quasar spectra whose \ion{C}{iv} EW $>70$\AA\,. Components "6" and "7" are correction components and are therefore permitted to have negative weights in the MCMC fitting.}
 \label{fig:COMP_high}
\end{figure*}

As discussed in Section \ref{sec:Methods}, we reconstruct spectra using Mean-Field Independent Component Analysis (MFICA). To account for the broad diversity of the \ion{C}{iv} emission line profiles, we require the use of three sets of MFICA components to reconstruct the quasar spectra. The objects whose \ion{C}{iv} EW $<20$\AA\, were fit using the components depicted in Fig. \ref{fig:COMP_low}, the objects whose $20<\,$\ion{C}{iv} EW$<70$\AA\, were fit using the components depicted in Fig. \ref{fig:COMP_int} and the remainder of the spectra were reconstructed using the components depicted in Fig. \ref{fig:COMP_high}. The \ion{C}{iv} EW thresholds were calculated by minimising the $\chi_{\nu}^2$ of the reconstructions, however, spectra in the overlapping regions of equivalent width generally have a similar quality fit with either set of MFICA components.

\section{Comparison to Wu \& Shen 2022}\label{Appendix:Shen}

Recently \citet{2022ApJS..263...42W} have published a catalogue of quasar continuum and emission line properties for the $\sim 750,000$ quasars in DR16Q, including updated systemic redshifts for the quasars. Here we explicitly compare our line properties to theirs. We match the 2,531 quasars at $3.5<z<4.0$ analysed in this paper to the \citet{2022ApJS..263...42W} catalogue by position and extract the appropriate rest-frame UV line properties for the comparison.

\begin{figure} 
 \includegraphics[width=\linewidth]{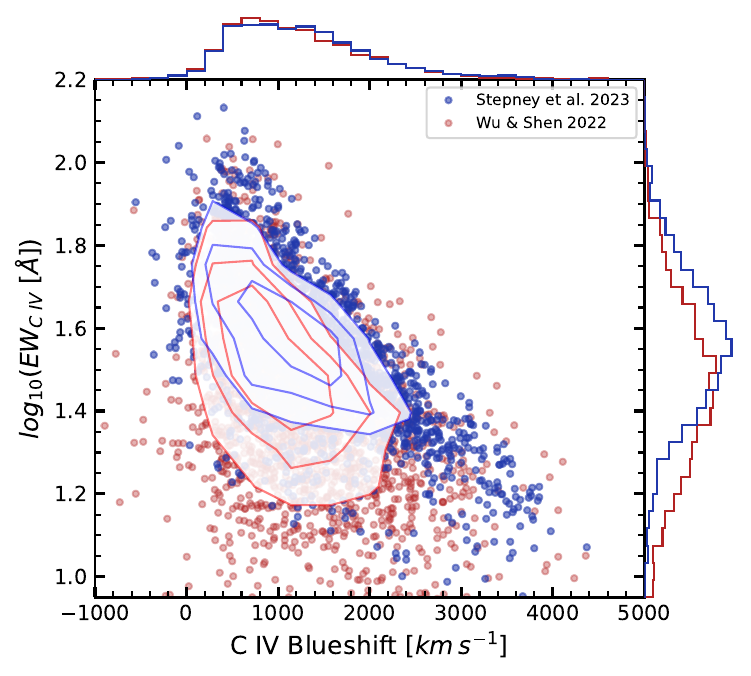}
 \caption{The \ion{C}{iv} emission space for the $3.5<z<4.0$ sample, using line properties derived from \citet{2022ApJS..263...42W} (red) and this work (blue). \ion{C}{iv} blueshifts are calculated from line centroids in both distributions. Density contours encircle  68, 50 and 25 per cent of the sample, respectively. Marginalised distributions of the \ion{C}{iv} blueshift and EW are also shown. The distributions do not feature the same characteristic shape in \ion{C}{iv} emission space, with the \citet{2022ApJS..263...42W} objects biased to lower \ion{C}{iv} EW.}
 \label{fig:Shen}
\end{figure}

Precision measurements of the \ion{C}{iv} emission line properties was not a primary aim of \citet{2022ApJS..263...42W}. Hence a composite Gaussian fitting recipe was adopted to model the emission line profile. The \ion{C}{iv} emission space for the $3.5<z<4.0$ sample is presented in Fig. \ref{fig:Shen}. While the marginalised distributions in \ion{C}{iv} blueshift are consistent with ours, the marginalised distributions in \ion{C}{iv} EW suggest that the \citet{2022ApJS..263...42W} catalogue is biased to lower \ion{C}{iv} EW when compared to this work. Since the low \ion{C}{iv} EW line profiles are poorly approximated by parametric fitting recipes, the characteristic tail to high blueshifts in the \ion{C}{iv} emission space is not as evident when one uses the \citet{2022ApJS..263...42W} measurements.

\begin{figure}
 \includegraphics[width=\linewidth]{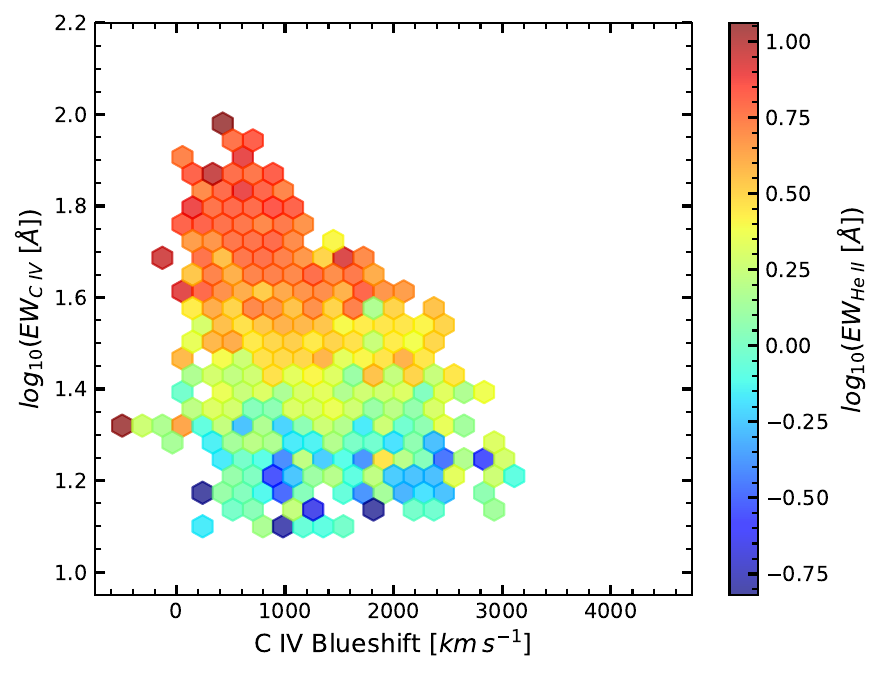}
 \caption{The median observed \ion{He}{ii} EW in bins of \ion{C}{iv} blueshift and \ion{C}{iv} EW for the $3.5<z<4.0$ sample, using line properties derived from \citet{2022ApJS..263...42W}. The \ion{He}{ii} EW is correlated with \ion{C}{iv} EW, however, the \ion{He}{ii} EW measurements are systematically overestimated when compared to this work. The correlation between the \ion{He}{ii} EW and the \ion{C}{iv} blueshift is also significantly weaker than what is observed in Fig. \ref{fig:CIV_HeII_hex}.}
 \label{fig:Shen_HEX}
\end{figure}

A key result of our paper is the correlation presented in Fig. \ref{fig:CIV_HeII_hex}. The existence of a correlation between the \ion{He}{ii} EW and both \ion{C}{iv} EW and blueshift could suggest that the \ion{C}{iv} emission is tracing outflows and that these are radiatively driven with an explicit dependence on the ionising SED. In Fig. \ref{fig:Shen_HEX} we present the same results using the \citet{2022ApJS..263...42W} measurements. In addition to the bias towards lower \ion{C}{iv} EW, the \ion{He}{ii} EW measurements appear systematically larger than those measured in this work. This difference can, in part, be attributed to the way in which the \ion{He}{ii} continuum is defined. Finally, while the correlation between \ion{He}{ii} EW and \ion{C}{iv} EW persists, in Fig. \ref{fig:Shen_HEX}, the anti-correlation between \ion{He}{ii} EW and \ion{C}{iv} blueshift is significantly weaker. This highlights the strength of the MFICA technique in recovering robust \ion{He}{ii} EWs and \ion{C}{iv} blueshifts for high-redshift quasar spectra. 

\section{Systemic redshift dependence on rest-frame wavelength range}\label{Appendix:Zsys}

\begin{figure}
 \includegraphics[width=\linewidth]{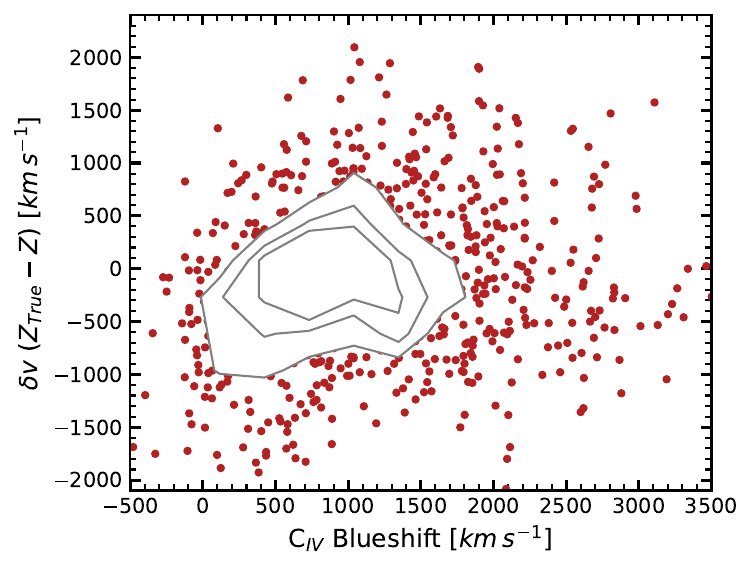}
 \caption{The velocity difference in systemic redshifts between those estimated by \citet{2020MNRAS.492.4553R} and calculated using the same recipe and rest-frame wavelength range as this work. This is plotted as a function of the \ion{C}{iv} blueshift for a random sample of 1000 $1.5<z<2.65$ quasars of comparable luminosity to the $3.5<z<4.0$ quasar population. At low \ion{C}{iv} blueshifts, the cross-correlation algorithm has a tendency to overestimate the systemic redshift and therefore biases the \ion{C}{iv} line centroids blue-ward of their true position.}
 \label{fig:Z2_systemic_testing}
\end{figure}

A key aim of this paper was to ensure that quasar samples across the entire redshift range have been analysed with the same methodologies for line-fitting and inference of quasar physical properties. However, at high-redshifts of $3.5<z<4.0$, the SDSS spectrum covers a more limited rest-frame wavelength range compared to quasars at $1.5<z<2.65$. In this paper we truncated the $3.5<z<4.0$ spectra at 2000\AA\ before estimating systemic redshifts and reconstructing the line profiles, due to the poor signal-to-noise quality towards the red-most end of the SDSS observed wavelength range. To assess the impact of the restricted wavelength range on systemic redshift estimates, we constructed a random sample of 1000 quasars from the UV luminosity matched $1.5<z<2.65$ quasars from \citet{2020MNRAS.492.4553R}. We then truncated the spectra at 2000\AA\, and ran our cross-correlation algorithm on these quasars to calculate systemic redshifts. These are then compared to the redshifts estimated by \citet{2020MNRAS.492.4553R} over a wider wavelength range in Fig. \ref{fig:Z2_systemic_testing}.

The redshift difference is systematic as a function of the \ion{C}{iv} blueshift below blueshifts of $\sim1000$ $\textnormal{km\,s}^{-1}$. The algorithm has a tendency to slightly overestimate systemic redshifts for objects whose \ion{C}{iv} blueshift $\leq 500 \textnormal{km\,s}^{-1}$. This is due to the presence of narrow associated absorbers predominantly affecting the \ion{C}{iv} lines at low blueshift but which are at low enough signal-to-noise to be not picked up by our narrow absorption masking procedure. As a result, the symmetric \ion{C}{iv} line profiles, with the majority of their flux at velocities close to the systemic redshift, are most affected. The implication for the results presented in Fig. \ref{fig:CIV-L3000_matched} and Fig. \ref{fig:CIV-MBH_LEDD_matched} is that since the algorithm more readily overestimates the systemic redshift at low \ion{C}{iv} blueshift, the position of the line centroid is shifted blue-ward as a result, hence the number of objects with \ion{C}{iv} blueshift $\leq 500 \textnormal{km\,s}^{-1}$ tails off more rapidly at $3.5<z<4.0$ compared to the lower redshift quasar population. The high \ion{C}{iv} blueshift tail on the other hand is less affected by this systemic redshift bias. The conclusions in this paper regarding the evolution of high \ion{C}{iv} blueshift quasars are therefore robust.  

%%%%%%%%%%%%%%%%%%%%%%%%%%%%%%%%%%%%%%%%%%%%%%%%%%

% Don't change these lines
\bsp	% typesetting comment
\label{lastpage}
\end{document}